\documentclass[preprint,5p,twocolumn]{elsarticle}
\usepackage{lipsum}
\biboptions{numbers,sort&compress} %citation
\usepackage{tikz} \usetikzlibrary{positioning} \tikzset{font=\footnotesize}
\usepackage{standalone}
\usepackage[numbers]{natbib}
\usepackage{graphicx} % Required for inserting images
\usepackage{svg}

\usepackage{color}
\usepackage{subcaption} % Required for subfigures
\usepackage{makecell}

\usepackage{amsmath,amssymb,amsfonts}

\usepackage[ruled, lined, linesnumbered, commentsnumbered, longend]{algorithm2e}
\usepackage{algpseudocode}

\usepackage{graphicx}
\usepackage{textcomp}
\usepackage{xcolor}
\usepackage{soul}
\usepackage{enumerate}
\usepackage{comment}

\usepackage{subcaption}
\usepackage{multirow}
\usepackage{hyperref}
\hypersetup{hidelinks}
\usepackage{tabu}
\usepackage{float}
\usepackage{booktabs}
\usepackage{arydshln}
\usepackage[inline]{enumitem}

\usepackage{amsmath}
\usepackage{mathtools}
\usepackage{textcomp}
\usepackage{xurl}
\usepackage{appendix}

\definecolor{boristext}{rgb}{0.22, 0.44, 0.33}
\definecolor{boriscomments}{rgb}{0.83, 0.0, 0.0}
\definecolor{davidcomments}{rgb}{0.0, 0.0, 0.83}
\definecolor{miguelcomments}{rgb}{0.5, 0, 0.8}

\begin{document}

\begin{frontmatter}
\title{Experimental Evaluation of Interactive Edge/Cloud Virtual Reality Gaming over Wi-Fi using Unity Render Streaming}

%\journal{arXiv}
\journal{Computer Communications}
\author{Miguel Casasnovas\corref{cor1}}
\ead{miguel.casasnovas@upf.edu}
\author{Costas Michaelides}
\ead{costas.michaelides@upf.edu}
\author{Marc Carrascosa-Zamacois}
\ead{marc.carrascosa@upf.edu}
\author{Boris Bellalta}
\ead{boris.bellalta@upf.edu}
\address{Wireless Networking Research Group, Universitat Pompeu Fabra \\ Carrer de Roc Boronat 138, 08018 Barcelona, Spain}

\cortext[cor1]{Corresponding author}
\begin{abstract}

Virtual Reality (VR) streaming enables end-users to seamlessly immerse themselves in interactive virtual environments using even low-end devices.
However, the quality of the VR experience heavily relies on Wireless Fidelity (Wi-Fi) performance, since it serves as the last hop in the network chain.
Our study delves into the intricate interplay between Wi-Fi and VR traffic, drawing upon empirical data and leveraging a Wi-Fi simulator.
In this work, we further evaluate Wi-Fi's suitability for VR streaming in terms of the Quality of Service (QoS) it provides.
In particular, we employ Unity Render Streaming to remotely stream real-time VR gaming content over Wi-Fi 6 using Web Real-Time Communication (WebRTC), considering a server physically located at the network's edge, near the end user.
Our findings demonstrate the system's sustained network performance, showcasing minimal round-trip time (RTT) and jitter at 60 and 90 frames per second (fps).
In addition, we uncover the characteristics and patterns of the generated traffic streams, unveiling a distinctive video transmission approach inherent to WebRTC-based services: the systematic packetization of video frames (VFs) and their transmission in discrete batches at regular intervals, regardless of the targeted frame rate.
This interval-based transmission strategy maintains consistent video packet delays across video frame rates but leads to increased Wi-Fi airtime consumption. Our results demonstrate that shortening the interval between batches is advantageous, as it enhances Wi-Fi efficiency and reduces delays in delivering complete frames.
\end{abstract}

\begin{keyword}
 Virtual Reality \sep Wi-Fi \sep Cloud Gaming \sep Edge Computing \sep Unity \sep WebRTC
\end{keyword}
\end{frontmatter}

%\maketitle

%\tableofcontents

%------------------------------------
%------------------------------------
%------------------------------------
%------------------------------------

\section{Introduction}

Virtual Reality (VR) has received significant attention for its transformative potential across multiple sectors, including multimedia, entertainment, gaming, healthcare, and education \cite{idris2020study}. Notably, the VR market is expected to witness significant growth, with an anticipated annual increase of 27.5\% from 2023 to 2030 \cite{website:grandviewresearch}. 
However, VR demands substantial computational power, particularly relying on high-performance Graphics Processing Units (GPUs) and Central Processing Units (CPUs) to achieve optimal rendering of graphics and maintain low-latency interactions for seamless, immersive gaming experiences.

In response to the computational demands of VR, remote rendering has emerged as a pivotal solution. By offloading resource-intensive rendering tasks to dedicated servers, it effectively alleviates the burden on local devices. Thereby, this approach not only enhances accessibility to sophisticated, high-quality VR games but also fosters the scalability of VR, accommodating a broader range of devices.

Remote rendering leverages distinct distributed computing paradigms, including cloud computing and edge computing.
Cloud computing relies on remote, centralized data centers for ubiquitous, on-demand access to shared resources. Hence, Cloud VR services such as PlutoSphere\footnote{\url{https://www.plutosphere.com/}} and XRStream\footnote{\url{https://xrstream.co/}} ---both powered by Nvidia CloudXR\footnote{\url{https://developer.nvidia.com/cloudxr-sdk/}}--- deliver VR experiences over the internet without requiring sophisticated local hardware.
Nevertheless, the inherent distance between users and cloud servers can lead to significant latency concerns.
In contrast, edge computing brings the computational resources to the network's edge, close to the end user. Thus, Edge VR minimizes latency and enhances real-time responsiveness, optimizing the delivery of immersive VR experiences.

Despite the benefits of remote rendering, delivering VR content over a network poses significant challenges, stemming primarily from the stringent bandwidth and latency requirements inherent to immersive VR experiences. Indeed, motion-to-photon latency should remain beneath 15~ms to prevent motion sickness~\cite{elbamby2018toward, alriksson2021xr}. Notably, ultimate VR is anticipated to demand substantial bandwidth, exceeding 1~Gbps, to achieve an unparalleled level of immersion~\cite{mangiante2017vr, huawei:cloudvrnetworksolution, bastug2017toward}.
In our previous research \cite{michaelides2023wi}, we empirically evaluated Edge VR streaming performance over Wireless Fidelity (Wi-Fi)~6 to multiple users using Air Light VR (ALVR)\footnote{\url{https://github.com/alvr-org/ALVR/}}, specifically in terms of latency. Our results demonstrated seamless VR streaming to three standalone Head-Mounted Displays (HMDs) at up to 100 Mbps per user, using either the Distributed Coordination Function (DCF) or DCF plus Orthogonal Frequency Division Multiple Access (OFDMA) only in the downlink (DL). Notably, uplink (UL) OFDMA scheduling disrupted the timely delivery of tracking data.

In this work, we leverage Unity’s Render Streaming plugin\footnote{\url{https://github.com/Unity-Technologies/UnityRenderStreaming/}} to deliver remotely rendered Edge VR gaming content over Wi-Fi~6 using Web Real-Time Communication (WebRTC)\footnote{\url{https://webrtc.org/}}. Our study delves into the characteristics of VR traffic, unveiling the presence of multiple periodic data streams with distinct generation patterns in both DL and UL directions.
Additionally, our work encompasses an experimental assessment of the performance of our VR streaming system, underscoring its ability to deliver VR content over Wi-Fi to end users, particularly at high frame rates.
We investigate Wi-Fi’s shaping effects on packet temporal arrangement, highlighting Access Point (AP) video packet aggregation and congestion-related delays.
Our video traffic analysis yields a significant finding: video frames (VFs) are segmented into packets that are transmitted in batches at regular intervals of 5.56~ms. This paced transmission strategy may be designed to enhance transmission efficiency by regulating the flow of video packets. Nevertheless, it prolongs the delivery of complete frames and leads to heightened Wi-Fi channel occupancy as higher frame rates are used.

Consequently, this work contributes to the field of VR streaming by:
\begin{enumerate}
    \item Showcasing the capability of our VR streaming system to deliver remotely rendered VR content using Wi-Fi. Indeed, VR Quality of Service (QoS) requirements are easily met under normal operating conditions, achieving low latency with Round-Trip Time (RTT) consistently below 5~ms, minimal jitter, and no packet loss at 60 and 90 frames per second (fps).
    \item Enhancing our understanding of WebRTC-based VR systems' traffic patterns, with an emphasis on the transmission of VFs arranged in temporally spaced batches of packets and the consequent impact on streaming performance indicators such as DL packet delays and VF assembly delays.
    \item Highlighting the interplay between Wi-Fi’s and Unity's remotely rendered VR traffic. In particular, we provide insights on the relationship between streaming parameters ---such as video frame rate and inter-batch time--- and Wi-Fi's packet aggregation capabilities. Additionally, we delve into their collective influence on spectrum resource utilization. 
\end{enumerate}

The paper is structured as follows: Section~\ref{sec:background} offers a comprehensive overview of the foundational elements of this work. Section~\ref{sec:related_work} overviews the related work. Section~\ref{sec:methodology} outlines our research methodology, including the experimental design. Sections \ref{sec:traffic_characteristics} to \ref{sec:wifi_vr_traffic_interplay} encompass the experimental evaluation. Section~\ref{sec:traffic_characteristics} details the traffic structure and data streams. Section~\ref{sec:video_traffic} delves deeper into the video stream. Section~\ref{sec:streaming_performance} thoroughly assesses streaming performance. Section~\ref{sec:wifi_vr_traffic_interplay} explores Wi-Fi’s and Unity’s VR traffic interplay. Finally, Section~\ref{sec:conclusions} concludes this work.

%------------------------------------
%------------------------------------
%------------------------------------
%------------------------------------
\section{WebRTC, Unity Render Streaming and Wi-Fi} \label{sec:background}

%------------------------------------
%------------------------------------
\begin{figure*}[t!!!]
  \centering
  \includegraphics[width=\textwidth]{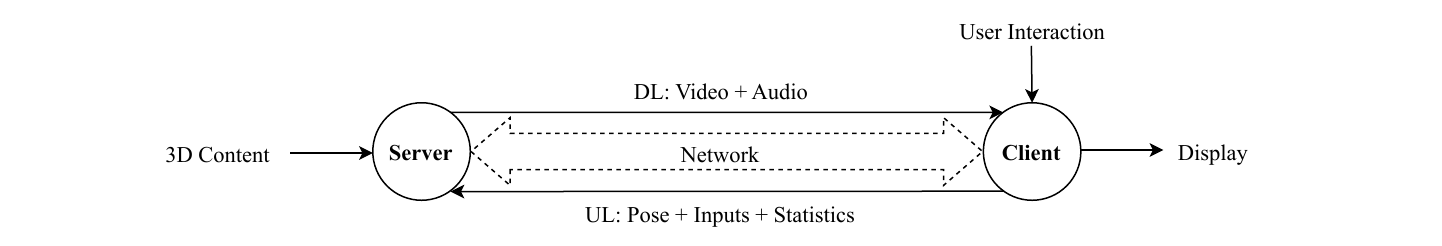}
  \caption{Remote streaming process. Adapted from \cite{shi2015survey}.}
  \label{fig:Remote_Streaming_Process}
\end{figure*}
%------------------------------------
%------------------------------------
\subsection{WebRTC}

WebRTC is an open-source project standardized by the World Wide Web Consortium (W3C)\footnote{\url{https://www.w3.org/}} and the Internet Engineering Task Force (IETF)\footnote{\url{https://www.ietf.org/}}. This technology enables secure peer-to-peer (P2P) communication, providing RTC capabilities via application programming interfaces (APIs) \cite{website:webrtc_spec}. 

WebRTC implementations incorporate congestion control mechanisms to enhance RTC efficiency. For instance, the Google Congestion Control (GCC) algorithm \cite{carlucci2016analysis}, implemented in commercial browsers such as Google Chrome, dynamically adjusts the transmission rate based on network conditions, using a delay-based controller at the receiver and a loss-based controller at the sender. Moreover, WebRTC integrates robust error resilience mechanisms to enhance transmission reliability, including Forward Error Correction (FEC) and Retransmission (RTX) \cite{holmer2013handling}. FEC entails the transmission of redundant information so that the receiver can correct errors without relying on retransmissions~\cite{website:RFC8854}. On the other hand, RTX involves the timely retransmission of lost packets in response to Negative Acknowledgments (NACKs) from the receiver~\cite{website:RFC4588}. Furthermore, WebRTC endpoints support several Real-Time Protocol (RTP) \cite{website:RFC8834, website:RFC3550} extensions, such as Picture Loss Indication (PLI) and Full Intra Request (FIR) \cite{website:RFC8834}. PLI messages request the sender to provide an intra-coded frame (I-frame) due to picture losses \cite{website:RFC4585}, while FIR messages demand an I-frame to avoid unusable video \cite{website:RFC5104}.

WebRTC relies on standardized protocols, including:
\begin{itemize}
    \item Interactive Connectivity Establishment (ICE)~\cite{website:RFC8445}: ICE is used to determine the optimal communication path. It uses STUN and TURN for Network Address Translation (NAT) traversal.
    \item Session Traversal Utilities for NAT (STUN)~\cite{website:RFC8489}: STUN is used to discover the public IP address and port of an endpoint located behind a NAT. Additionally, it is used to assess connectivity and serves as a keep-alive mechanism to maintain NAT bindings. 
    \item Traversal Using Relays around NAT (TURN)~\cite{website:RFC8656}: TURN is used to relay data when direct P2P communication is hindered by network constraints.
    \item Session Description Protocol (SDP)~\cite{website:RFC8866}: SDP is used to exchange media capabilities and session details. 
    \item Secure Real-Time Protocol (SRTP)~\cite{website:RFC3711}: SRTP ensures the secure transport of real-time media, providing confidentiality, integrity, and authentication.
    \item Secure Real-Time Control Protocol (SRTCP)~\cite{website:RFC3711}: SRTCP securely delivers control and statistical information related to the RTP media stream. 
    \item Datagram Transport Layer Security (DTLS)~\cite{website:RFC6347, website:RFC5764}: DTLS provides security for datagram-based communications through integrated key management and parameter negotiation.
\end{itemize}

%------------------------------------
%------------------------------------

\subsection{Unity Render Streaming}
\label{subsec:unity_render_streaming}
Unity\footnote{\url{https://unity.com/}} is a versatile cross-platform game engine designed for creating interactive 2D and 3D experiences. Unity uses C\# as its primary programming language and integrates drag-and-drop functionalities. Additionally, it provides built-in features and developer tools specifically tailored for building immersive Extended Reality (XR)\footnote{Extended Reality encompasses Virtual Reality (VR), Augmented Reality (AR), and Mixed Reality (MR)} experiences, including support for the OpenXR\footnote{\url{https://www.khronos.org/openxr/}} standard.

Unity’s Render Streaming plugin enables remote real-time rendering and streaming of interactive experiences to the web using WebRTC. This open-source solution offloads resource-intensive rendering processes from the client’s device to a dedicated rendering server running Unity.

Signaling is handled by a \emph{node.js} web server, enabling direct P2P communication via User Datagram Protocol (UDP)~\cite{website:RFC768}. The rendering server integrates Unity’s WebRTC package\footnote{\url{https://github.com/Unity-Technologies/com.unity.webrtc/}}, while the client relies on WebRTC’s native JavaScript APIs within the web browser. Nevertheless, both peers leverage common interfaces, including \emph{RTCPeerConnection} to establish and manage the P2P connection and \emph{RTCDataChannel} for reliable bidirectional communication of non-media data.

Unity Render Streaming enables the selection of the video codec, the target frame rate (default up to 60 fps), the maximum and minimum video streaming bitrates (default up to 10 Mbps), and the resolution. Nevertheless, it does not provide options for configuring codec-specific properties. Notably, Unity’s WebRTC package uses an infinite Group of Pictures (GOP) length and intra-refresh~\cite{nvidia_video_codec_api}. Thus, in Unity Render Streaming, there are no periodic I-frame transmissions.

Unity Render Streaming offers support for several input methods within the web browser, including mouse, keyboard, touch, and gamepad. It does not currently offer support for VR specific systems as remote input sources. Nevertheless, it can be seamlessly integrated with WebXR\footnote{\url{https://immersiveweb.dev/}} to capture the input from VR devices, as demonstrated in~\cite{seligmann2020web, githubFusedVR}.

The remote streaming process using Unity’s Render Streaming plugin unfolds as depicted in Fig.~\ref{fig:Remote_Streaming_Process}. In particular, the client captures user interactions within the web browser and sends them to the server. The server processes the incoming data, updates the game state, and generates the corresponding visual content. The rendered content is encoded into a video stream and transmitted to the client. Upon reception, the client decodes and displays the visual content in the user’s web browser.

%------------------------------------
%------------------------------------
\subsection{Wi-Fi 6}

Wi-Fi~6 (IEEE 802.11ax) is an amendment to the IEEE 802.11 Physical~(PHY) and Medium Access Control~(MAC) layers to enhance operational efficiency~\cite{80211ax2021amendment, bellalta2016ieee, khorov2018tutorial}.  

\begin{figure}[ht!]
  \centering
    \includegraphics[width=\columnwidth]{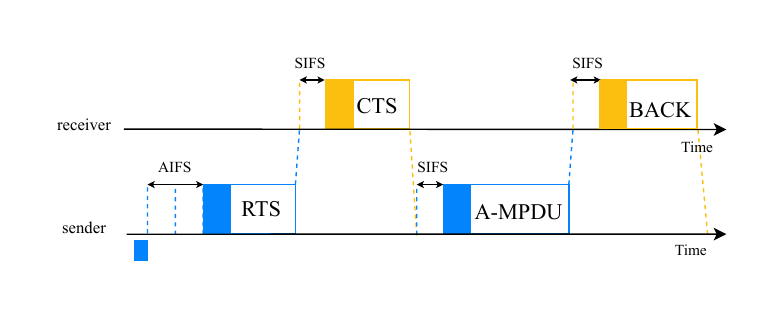}
  \caption{SU EDCA: CSMA/CA Process.}
  \label{fig:DIFS_SIFS_process}
\end{figure}

Wi-Fi~6 operates on 2.4 and 5~GHz frequency bands, supporting channel bandwidths up to 160~MHz. Wi-Fi~6E extends this standard into the 6 GHz spectrum. The default channel access scheme of Wi-Fi~6 is the Enhanced Distributed Channel Access (EDCA).
EDCA relies on Carrier Sense Multiple Access with Collision Avoidance~(CSMA/CA) and optional Request to Send (RTS) and Clear to Send (CTS) messages to manage access to the communication medium. In the presence of multiple distinct flows, EDCA provides traffic differentiation capabilities by categorizing traffic into four Access Categories: Voice, Video, Best Effort, and Background. The basic units of data transmission are MAC Protocol Data Units (MPDUs). However, multiple MPDUs can be aggregated into a single Aggregated MPDU (A-MPDU) to enhance data transmission efficiency by reducing overhead. This process is commonly referred to as \emph{packet aggregation}.
Fig.~\ref{fig:DIFS_SIFS_process} describes the basic Single User (SU) channel access procedure, incorporating the RTS/CTS mechanism. The sender monitors the channel during an Arbitration InterFrame Space (AIFS) interval. If the channel remains idle throughout this interval, the sender initiates a RTS message. Upon reception, the receiver replies with a CTS message after a Short Interframe Space (SIFS). Subsequently, the sender transmits an A-MPDU after another SIFS. Upon successful data reception, the receiver sends a Block acknowledgement (BACK) after a SIFS, confirming the transmission's success. If the BACK is not received within a certain timeout period, the sender initiates a retransmission of all transmitted packets. Otherwise, upon receiving a BACK, the sender identifies the MPDUs that were not received successfully ---if any--- and proceeds to retransmit them in the subsequent transmission.

Moreover, Wi-Fi~6 introduces OFDMA for frequency multiplexing (F-MUX). OFDMA divides the available frequency spectrum into multiple orthogonal sub-channels, called Resource Units (RUs), enabling multiple users to access the network simultaneously. Wi-Fi~6 also integrates both SU and Multi-User (MU) Multiple Input Multiple Output (MIMO) for spatial multiplexing (S-MUX), incorporating MU-MIMO in the UL. MIMO enables the simultaneous transmission and reception of multiple data streams by using multiple antennas.

This standard offers substantial improvements in data rates, network capacity, and efficiency compared to its predecessor, Wi-Fi~5 (IEEE 802.11ac) ---please refer to Table~\ref{tab:wifi6wifi5} for a comparative overview. Hence, Wi-Fi~6 stands out as a compelling candidate for providing the essential connectivity required to enable real-time, wireless VR gaming experiences. %For more information on Wi-Fi 6 see \cite{sankaran2021wi}. 

%Table Wi-Fi 6 vs Wi-Fi 5
\begin{table}[ht!]
\centering
\caption{Wi-Fi 5 and Wi-Fi 6 features comparison. Adapted from ~\cite{oughton2021revisiting}.}
\label{tab:wifi6wifi5}
\footnotesize
\begin{tabular}{@{}lll@{}}
\toprule
\textbf{} & \textbf{Wi-Fi 5}   & \textbf{Wi-Fi 6} \\                      
\textbf{Frequency} & 5 GHz & 2.4, 5, 6 GHz           \\
\textbf{Bandwidth (MHz)} & 20,40,80,80+80,160   & 20,40,80,80+80,160 \\  
\textbf{MCS} & up to 256-QAM    & up to 1024-QAM \\   
\textbf{Data Rate} & up to 7 Gbps & up to 9.6 Gbps \\
\textbf{Target Wake Time} & No   & Yes \\   
\textbf{BSS Coloring} & No   & Yes \\
\textbf{F-MUX} & OFDM & OFDM, \\  
\textbf{} &  & DL/UL OFDMA   \\
\textbf{S-MUX} & SU-MIMO,  & SU-MIMO, \\   
\textbf{} & DL MU-MIMO  &  DL/UL MU-MIMO   \\
 \bottomrule
\end{tabular}
\end{table}

%------------------------------------
%------------------------------------
%------------------------------------
%------------------------------------

\section{Related work}\label{sec:related_work}

In the landscape of remote gaming, Cloud Gaming (CG) has garnered significant research attention, leading to studies on its protocols and traffic dynamics. 
For instance, Di Domenico et al.~\cite{di2021network} identified distinct protocol implementations in several CG services. Notably, Google Stadia\footnote{Shutdown in January 2023, \url{https://stadia.google.com/gg/}} leveraged WebRTC, GeForce NOW\footnote{\url{https://play.geforcenow.com/}} relied on RTP for multimedia streaming without standardized session establishment protocols, and PlayStation Now\footnote{Recently integrated to PlayStation Plus, \url{https://www.playstation.com/en-us/ps-plus/}} adopted a proprietary UDP-based approach. 
Moreover, multiple studies have modeled the traffic generated by CG platforms using client-side gathered data~\cite{manzano2014dissecting, carrascosa2022cloud}.
Manzano et al.~\cite{manzano2014dissecting} meticulously examined the OnLive\footnote{Shutdown in April 2015} CG platform traffic, unveiling the characteristics of multiple RTP streams. In particular, the RTP video flow exhibited regular VF bursts, split into multiple packets. Interestingly, video packets demonstrated inter-arrival times of up to 6~ms. 
Carrascosa et al.~\cite{carrascosa2022cloud} conducted an extensive assessment of Google Stadia’s traffic, uncovering traffic generation patterns within the downstream RTP stream: a consistent $1/\text{fps}$ inter-frame interval and $\approx 2$~ms interval video batches within each frame period.

Nevertheless, research on cloud-driven VR is scarce. 
Notably, Zhao et al.~\cite{zhao2021virtual} investigated VR CG under fixed and adaptive bitrate encoding schemes, using Paperspace\footnote{\url{https://www.paperspace.com/}}, a commercial CG service provider, and Virtual Desktop\footnote{\url{https://www.vrdesktop.net/}}. The authors highlighted the advantages of adaptive bitrate, including reduced latency and frame loss. Additionally, their traffic analysis disclosed two successive sequences of video packets within each VF, providing distinct perspectives for each eye.

In contrast, several works have replicated Cloud VR schemes within controlled, local environments using edge computing servers. %\cite{korneev2022studying, li2020performance, vikberg2021optimizing, lee2021measurement}
For instance, Li et al.~\cite{li2020performance} evaluated ALVR’s performance, highlighting the substantial impact of limited bandwidth and high packet loss on frame rate and latency. Surprisingly, VR gamers barely perceived RTT values up to 90~ms. 
Vikberg’s experimental research~\cite{vikberg2021optimizing} focused on optimizing and evaluating WebRTC for Cloud VR using Unity Render Streaming. The author analyzed diverse configurations to minimize latency, pinpointing resolution reduction and rendering rate increase as effective strategies.
Lee et al.~\cite{lee2021measurement} leveraged Unity’s WebRTC-based remote streaming solution, showing that an increase in concurrent users reduced each user’s data rate and led to increased jitter and frame drops.
On the other hand, Korneev et al. \cite{korneev2022studying, korneev2024model} analyzed stereoscopic RTP VR traffic using the Pico Streaming Assistant\footnote{\url{https://www.picoxr.com/global/software/pico-link/}}, highlighting single-slice VF encoding and bimodal frame size distribution. Both studies modeled the traffic, but \cite{korneev2024model} considered additional VR application peculiarities, including inter-frame time inconsistencies and stereo video streams.

Similarly, numerous studies have leveraged servers close to end-users for wireless VR streaming over Wi-Fi. %\cite{ salehi2020traffic,  lecci2021open, chiariotti2023temporal, alhilal2023network,  jansen2023can}
Salehi et al.~\cite{salehi2020traffic} explored edge-enabled VR traffic using ALVR, uncovering that VFs are generated every $1/\text{fps}$ and fragmented into multiple MPDUs. Their findings indicated that increasing bitrate does not consistently enhance quality across applications and results in larger VFs that may impose additional strain on the network. % Additionally, their results indicated that the H.265 codec outperforms H.264, producing higher-quality videos with reduced file sizes.
Building upon their work in~\cite{lecci2021open}, Chiariotti et al. \cite{chiariotti2023temporal} conducted a fundamental VR traffic characterization using RiftCat 2.0\footnote{\url{https://riftcat.com/vridge/}}. The authors unveiled the underlying traffic streams and highlighted the occurrence of frame rate aligned VF bursts. Interestingly, the packets comprising a VF were transmitted collectively within a single batch.
Alhilal et al.~\cite{alhilal2023network} examined the impact of concurrent users on network data transfer rates within social WebRTC-based VR platforms such as Mozilla Hubs\footnote{\url{https://github.com/mozilla/hubs/}}. The authors observed that an increase in the number of concurrent users led to a substantial rise in startup delay.
Jansen et al.~\cite{jansen2023can} assessed the network performance and requirements for wireless VR offloading to the edge using Android Debug Bridge\footnote{\url{https://developer.android.com/studio/command-line/adb/}}. Their findings underscored the need for a stable, high-throughput network. Interestingly, despite lower resource utilization, Edge VR led to increased battery depletion compared to native processing.

Thus, research on VR streaming has predominantly revolved around performance~\cite{li2020performance, vikberg2021optimizing, lee2021measurement, jansen2023can} and traffic characteristics~\cite{zhao2021virtual, korneev2022studying, korneev2024model, salehi2020traffic,lecci2021open, chiariotti2023temporal, alhilal2023network}, using diverse hardware such as standalone HMDs~\cite{zhao2021virtual, salehi2020traffic, alhilal2023network, korneev2022studying, korneev2024model, li2020performance, vikberg2021optimizing}, smartphones~\cite{lecci2021open, chiariotti2023temporal, vikberg2021optimizing}, and computers~\cite{lee2021measurement}. Notably, several studies have further contributed to the field by presenting VR traffic models~\cite{korneev2022studying, korneev2024model, lecci2021open, chiariotti2023temporal}. 

To the best of our knowledge, this is the first work in the literature that delves into the intricacies of Unity’s WebRTC-based VR traffic and the transmission of VFs in temporally spaced batches of packets. Indeed, prior studies in CG and VR remote rendering highlighted distinct RTP video traffic dynamics~\cite{carrascosa2022cloud,korneev2022studying, korneev2024model, lecci2021open, chiariotti2023temporal,zhao2021virtual}.
Our study not only demonstrates the system's ability to meet VR QoS requirements but also delves into the influence of WebRTC's video transmission approach on streaming performance, Wi-Fi's packet aggregation, and spectrum efficiency.

%------------------------------------
%------------------------------------
%------------------------------------
%------------------------------------

\section{Methodology} \label{sec:methodology}

In this section, we provide a detailed description of the experimental environment and the data collection and analysis procedures.

%------------------------------------
%------------------------------------
\subsection{Experimental setup}
\label{sec:experimental_setup}
%------------------------------------
%\subsubsection{Hardware}

\begin{figure}[ht]
  \centering
\includegraphics[ width=0.65\columnwidth]{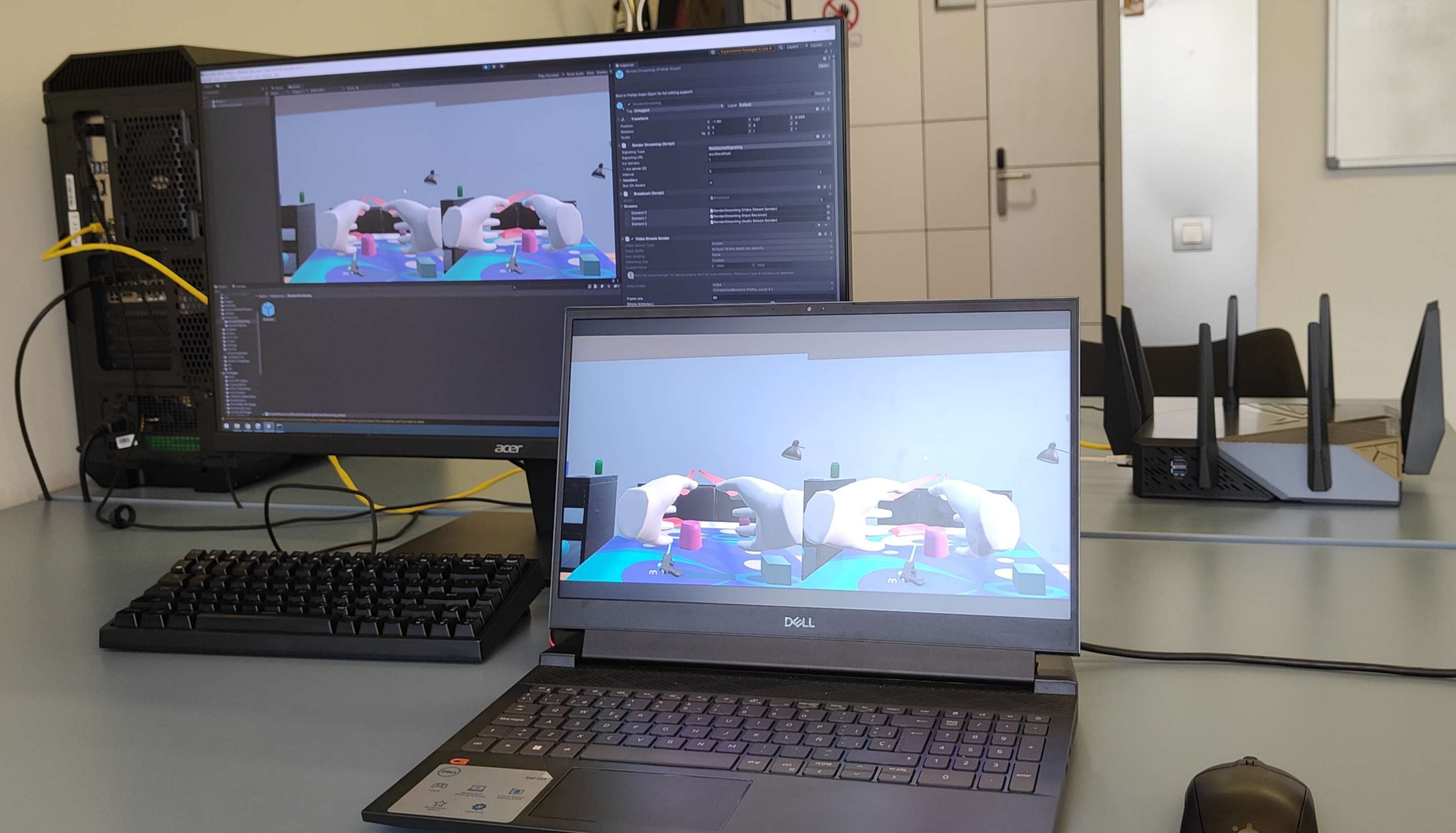}
  \caption{A snapshot of the experimental setup.}
  \label{fig:Setup_snapshot}
\end{figure}

The experiments were conducted in a controlled local environment featuring a high-performance gaming desktop acting as the streaming server, a gaming-centric AP, and a VR-Ready laptop serving as the client, as shown in Fig.~\ref{fig:Setup_snapshot}.

\begin{figure}[ht]
  \centering
\includegraphics[width=\columnwidth]{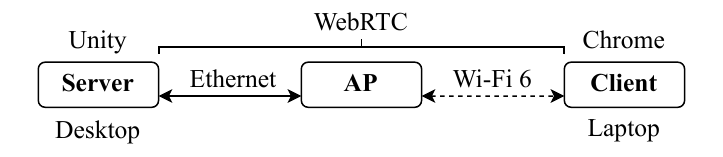}
  \caption{Testbed components.}
  \label{fig:Testbed_Components}
\end{figure}

The testbed was established at the UPF’s Department of Information and Communication Technologies (DTIC) in Barcelona, Catalonia, Spain. The AP is Wi-Fi~6 compliant and works at the 5 GHz band using a 80~MHz channel.
As illustrated in Fig.~\ref{fig:Testbed_Components}, the server was connected to the AP via Ethernet, ensuring a stable and reliable connection. 
The client sustained a robust Wi-Fi~6 connection with a Received Signal Strength Indicator (RSSI) of $-46$~dBm, leveraging two spatial streams (SS) for enhanced data transmission. The favorable network conditions enabled the utilization of the highest achievable modulation and coding scheme (MCS): 1024-QAM and a 5/6 coding rate.
Thus, a theoretical network capacity of up to 1200~Mbps, considering an 80~MHz channel, 2~SS, MCS~11, and an 800~ns Guard Interval. However, measurements conducted using iperf\footnote{\url{https://iperf.fr/}} indicated a capacity of 600~Mbps in average.
Equipment specifications are detailed in Table~\ref{tab:Experimental_equipment}. 
%------------------------------------
%------------------------------------
%Table experimental equipment
\begin{table}[ht]
\centering
\caption{Experimental equipment.}
\label{tab:Experimental_equipment}
\footnotesize
\begin{tabular}{@{}lllll@{}}
\toprule
\textbf{Desktop} & CPU   & i5-12600KF \\                      
\textbf{} & SSD & 500GB NvMe            \\
\textbf{} & GPU   & NVIDIA\textsuperscript{\tiny\textregistered} GeForce RTX™ 3080, 10 GB \\                      
\textbf{} & NIC & Realtek Gaming 2.5GbE \\
\textbf{} & RAM   & 32GB DDR5 4800 MHz \\                      
\textbf{} & OS  & Windows 10 x64 \\
\textbf{AP} & Model & ASUS ROG Rapture GT-AX11000 \\   
\textbf{} & FW  & 3.0.0.4.388\_22525 \\
\textbf{Laptop}  & Model & Dell G15 5521 Special Edition \\                   
\textbf{} & CPU & i7-12700H \\              
\textbf{} 
& GPU   & NVIDIA\textsuperscript{\tiny\textregistered} GeForce RTX™ 3060 Laptop \\ 
\textbf{} 
& WNIC &
Intel\textsuperscript{\tiny\textregistered} Killer™ Wi-Fi 6 AX1650i 2x2 \\
\textbf{} & OS  & Windows 11 x64 \\ 
\bottomrule
\end{tabular}
\end{table}

\begin{figure*}[]
  \centering
  \begin{subfigure}[b]{0.3\linewidth}
    \includegraphics[width=\linewidth]{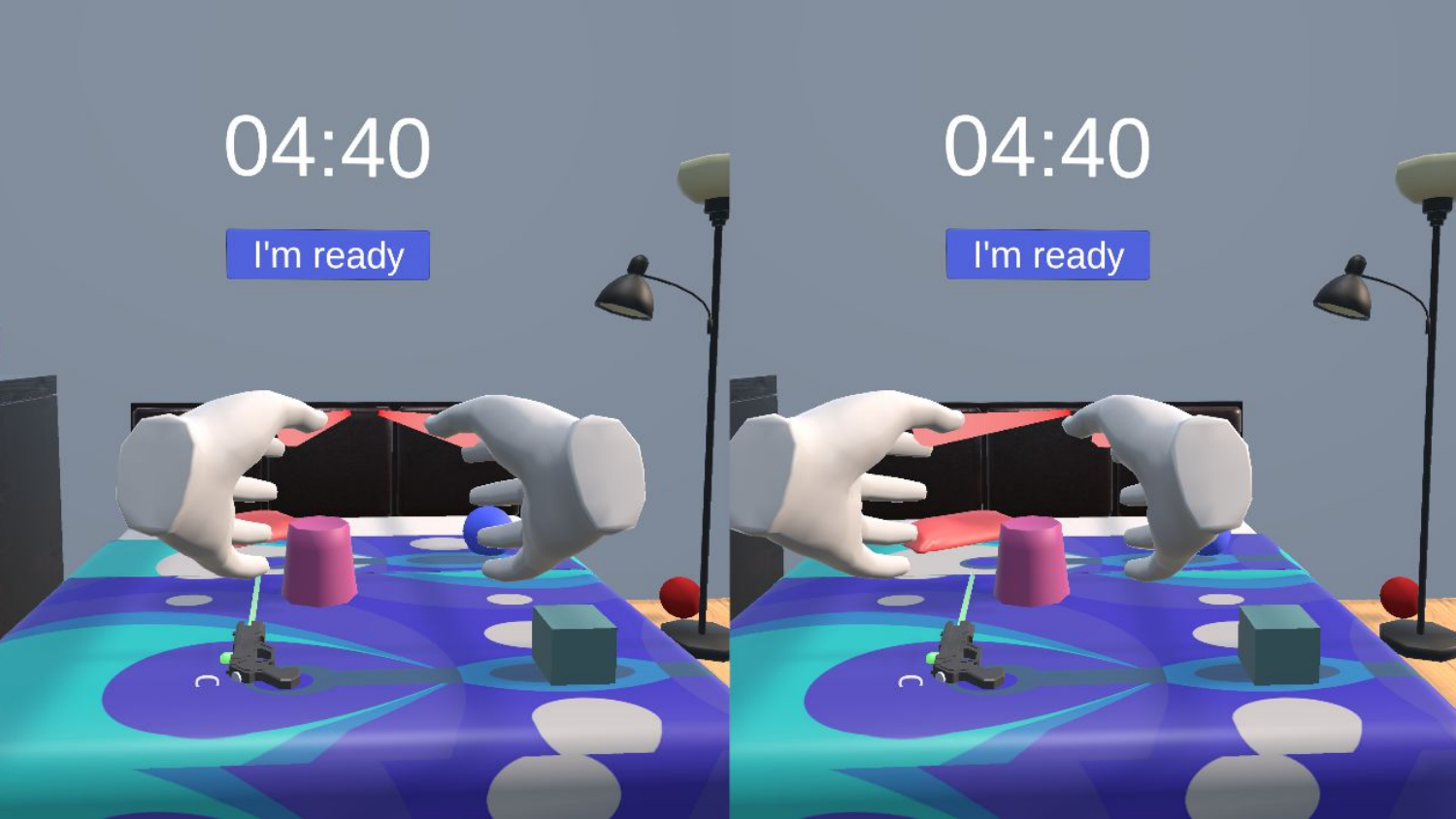}
    \caption{Alteration Hunting (AH)}
    \label{fig:AH_Display}
  \end{subfigure}
  \begin{subfigure}[b]{0.3\linewidth}
    \includegraphics[width=\linewidth]{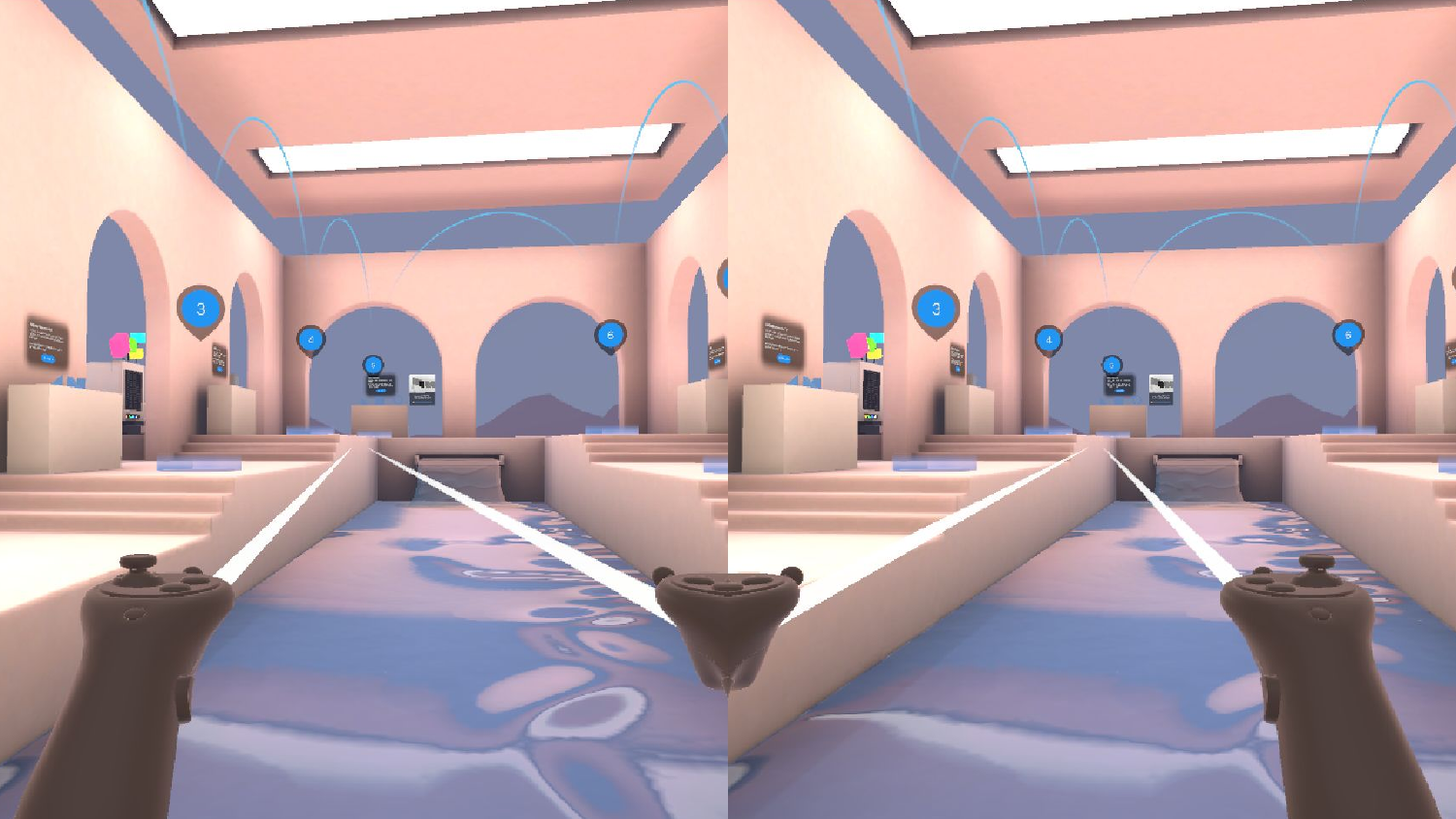}
    \caption{XR Interaction Toolkit demo (ITK)}
    \label{fig:ITK_Display}
  \end{subfigure}
  \begin{subfigure}[b]{0.3\linewidth}
    \includegraphics[width=\linewidth]{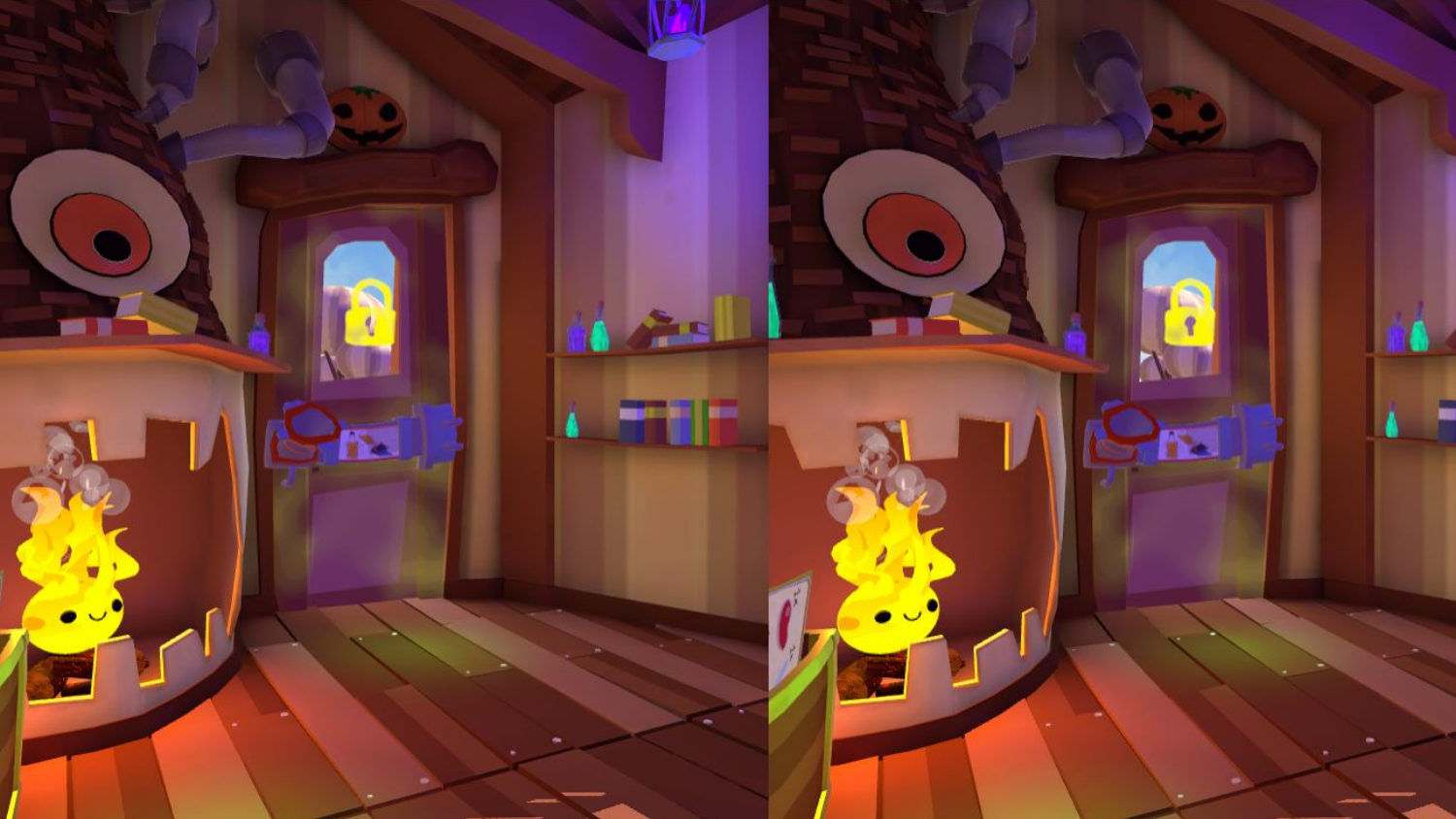}
    \caption{VR Beginner: The Escape Room (ER)}
    \label{fig:ER_Display}
  \end{subfigure}
  \caption{Screenshots of the VR games.}
  \label{fig:applications_screenshot}
\end{figure*}
%------------------------------------
%------------------------------------
%\subsubsection{Software}
The server, running Unity (v2021.3.6f1), rendered and streamed the VR games content to the client over WebRTC\footnote{Based on the WebRTC SDP information, in-band FEC was enabled only for audio, and RTX, PLI, and FIR were enabled for video.
%to ensure robust error resilience during the streaming process. 
Nevertheless, during our tests, neither audio FEC packets nor video FIR messages were received. PLI messages (and their associated I-frames) and RTX requests were minimal.} using Unity Render Streaming. It also served as the signaling server through a WebSocket connection.  
The client accessed the web-based application using the Google Chrome browser (v113.0.5672.93) and decoded and displayed the rendered content while capturing user inputs. Using a laptop enabled capturing network traffic on the client side, seamlessly integrating keyboard and mouse input methods supported by Unity's Render Streaming. Pulover’s Macro Creator\footnote{\url{https://www.macrocreator.com/}} (v5.4.1) facilitated the recording and playback of predefined mouse and keyboard inputs at the client, ensuring a systematic and reproducible testing environment. This method enabled continuous adjustment of the viewpoint's rotation, reproducing the sensitivity of VR HMDs in tracking subtle changes in head orientation. Consequently, as outlined in \ref{sec:HMDcomparative}, the video generated using either a laptop or an HMD as input sources remained consistent in characteristics. 

The foundation of our investigation relied on \emph{Alteration Hunting} (AH)\footnote{\url{https://github.com/miguelcUPF/AlterationHunting/}}, an immersive VR game we developed using Unity v2021.3.6f1. In addition, we conducted a comparison of AH’s traffic characteristics with two open-source sample scenes: the \emph{XR Interaction Toolkit 2.3} (ITK)\footnote{\url{https://github.com/Unity-Technologies/XR-Interaction-Toolkit-Examples/}} demo scene and \emph{The Escape Room} (ER)\footnote{\url{https://assetstore.unity.com/packages/templates/tutorials/vr-beginner-the-escape-room-163264/}} VR experience from Unity Learn. Each game was customized to integrate Unity’s remote streaming solution, including Unity’s Render Streaming plugin (v3.1.0-exp.4) and Unity’s WebRTC package (v3.0.0-pre.1). Unity's Render Streaming plugin was extended to support target frame rates of up to 90~fps and streaming bitrates of up to 100~Mbps, surpassing the constraints of 60~fps and 10~Mbps in the \textit{VideoStreamSender.cs} script, as outlined in Section~\ref{subsec:unity_render_streaming}.
Furthermore, the VR games were upgraded to present a stereoscopic view by placing left and right eye cameras with a 63~mm interpupillary distance, ensuring that each camera spans half of the viewport horizontally, as depicted in Fig.~\ref{fig:applications_screenshot}.

Notably, the video content was encoded in the H.264 Constrained Baseline 5.1 format using the NVENC\footnote{\url{https://developer.nvidia.com/video-codec-sdk/}} hardware-accelerated encoder. The video streaming resolution was set to 3664x1920p (1832x1920p per eye), adhering to the specifications of several HMDs, such as the Oculus Quest~2.
Additionally, the experiments encompassed various target frame rates, including 90, 60, and 30~fps, to assess their influence on system performance and traffic dynamics. To ensure consistent network load levels and facilitate the isolation of the target frame rate's influence on traffic, both the minimum and maximum video streaming bitrates were set at 50~Mbps, maintaining a constant bitrate.
Nevertheless, \ref{sec:CBRvsDBR} investigates the traffic characteristics and streaming performance when the minimum and maximum bitrate parameters are different, thus enabling dynamic bitrate.

%Table Conf details
\begin{comment}
    \begin{table}[h]
\centering
\caption{Streaming and Network Settings }
\label{tab:Configuration_details}
\footnotesize
\begin{tabular}{@{}lllll@{}}
\toprule
\textbf{Unity} & Codec & H.264 \\
\textbf{} & Resolution   & 3664x1920p \\ 
\textbf{} & Bitrate   & 50 Mbps (constant) \\   
\textbf{} & Frame rate & 30 fps, 60 fps, 90 fps  \\
\textbf{AP} & Frequency & 5 GHz\\                      
\textbf{} & Bandwidth  & 80 MHz \\
\textbf{} & Standard  & 802.11ax \\
\textbf{}  & MCS & Up to MCS 11 (1024-QAM) \\                    
 \bottomrule
\end{tabular}
\end{table}
\end{comment}

\subsection{Data collection and analysis}

Data collection encompassed multiple one-minute trials, including three for AH at each target frame rate (90, 60, and 30~fps) and one each for ITK and ER at 90~fps. Wireshark (v4.0.5)\footnote{\url{https://www.wireshark.org/}} was used for capturing network traffic from both endpoints. The collected traffic traces were processed using \emph{tshark} to extract relevant UDP data into a CSV file.
WebRTC statistics~\cite{website:webrtc_stats}, encompassing encoded and decoded fps\footnote{\textit{framesEncoded} field from \textit{RTCOutboundRtpStreamStats} and \textit{framesDecoded} field from \textit{RTCInboundRtpStreamStats}}, retransmissions\footnote{\textit{retransmittedPacketsSent} field from \textit{RTCOutboundRtpStreamStats}}, RTT\footnote{\textit{roundTripTime} field from \textit{RTCRemoteInboundRtpStreamStats}}, jitter\footnote{\textit{jitter} field from \textit{RTCReceivedRtpStreamStats}}, and packet loss\footnote{\textit{packetsLost} field from \textit{RTCReceivedRtpStreamStats}} were also gathered from the client and server. Server statistics were extracted in JSON format using Unity’s WebRTC package, whereas client statistics were stored in TXT format via \url{chrome://webrtc-internals}. 
Hence, each dataset\footnote{The datasets are publicly available on Zenodo \cite{datasetvr2024}} includes tshark-processed traffic traces and WebRTC statistics from both the client and server.

Data analysis, leveraging the Python programming language and the Pandas library\footnote{\url{https://pandas.pydata.org/}}, focused on a 30-second window of active gameplay.

%------------------------------------
%------------------------------------
%------------------------------------
%------------------------------------
\section{Traffic characteristics}\label{sec:traffic_characteristics}

In this section, our focus is on analyzing the traffic structure and identifying the data streams involved in the transmission process. The analysis relies on traces collected from the server side for DL traffic and from the client side for UL traffic, providing insights into traffic generation in both directions. It encompasses Alteration Hunting, the XR Interaction Toolkit demo, and Escape Room at 90~fps.

%------------------------------------
%------------------------------------
\subsection{Overview}

\begin{figure}[tt!]
  \centering
  \includegraphics[width=\columnwidth]{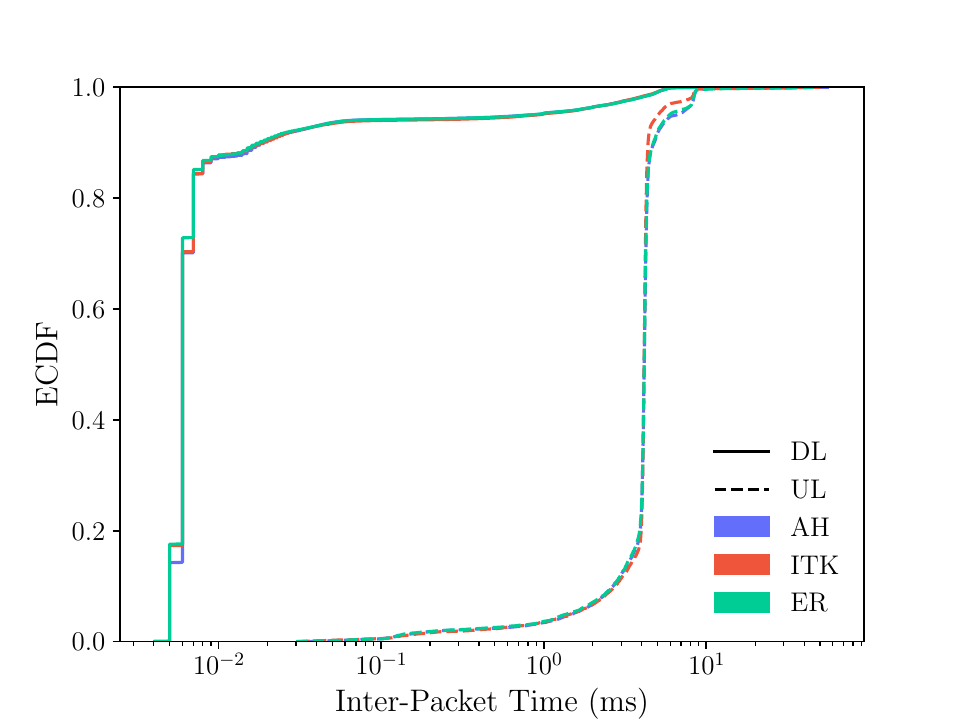}
  \caption{Inter-packet time ECDF for different games at 90 fps.}
  \label{fig:Inter_Packet_Time_ECDF_Different_Games}
\end{figure}

Fig.~\ref{fig:Inter_Packet_Time_ECDF_Different_Games} displays the empirical cumulative distribution function (ECDF) of the the inter-packet time for both DL and UL directions, showing consistency among the analyzed games. Remarkably, the DL consistently exhibits shorter time intervals between packets, indicating a higher packet transfer rate. Indeed, over 95.8\% of the data packets are transmitted in the DL, underscoring a significant disparity in packet distribution.

\subsection{Traffic streams}

Several WebRTC data streams, leveraging UDP as the underlying transport protocol, have been observed, each serving distinct purposes:

\begin{itemize}
    \item STUN messages are exchanged to check connectivity between endpoints.
    \item The SRTP stream securely transports real-time audio and video data, distinguishable by payload type or Synchronization Source Identifier (SSRC).
    \item The SRTCP stream carries control information, providing feedback on the transmitted data quality.
    \item The DTLS stream is responsible for delivering arbitrary data through the data channel, including input and pose data in the UL.
    \item Generic UDP packets are potentially related to network synchronization or timing.
\end{itemize}

\begin{table*}[t!!]
\centering
\caption{Traffic streams characteristics for different games at 90 fps: mean packet size (Bytes), mean inter-packet time (ms), and load (Mbps).}
\label{tab:traffic_streams_characteristics}
\footnotesize
\begin{tabular}{@{}llllllllll@{}}
\multirow{2}{*}{\textbf{}} & \multicolumn{3}{l}{\textbf{AH}} & \multicolumn{3}{l}{\textbf{ITK}} & \multicolumn{3}{l}{\textbf{ER}} \\
\cmidrule(lr){2-4} \cmidrule(lr){5-7} \cmidrule(lr){8-10}
& Bytes & ms & Mbps & Bytes & ms & Mbps &Bytes & ms & Mbps \\
\midrule
\textbf{Downlink} &&&&&&&&& \\
STUN & 122 & 1287.42 & 0.0008 & 122 & 1288.10 & 0.0008 & 122 & 1288.45 & 0.0008 \\
SRTP Audio & 83 & 20 & 0.033 & 83 & 20.01 & 0.033 & 730.90 & 20.01 & 0.29 \\
SRTP Video & 1242.16 & 0.19 & 53.45 & 1242.61 & 0.19 & 53.63 & 1243.56 & 0.19 & 53.38 \\
DTLS & 107.01 & 8.94 & 0.096 & 107.01 & 8.74 & 0.098 & 107.01 & 8.89 & 0.096 \\
Generic UDP & 112 & 8.26 & 0.11 & 112 & 8.24 & 0.11 & 112.16 & 8.23 & 0.11 \\
\midrule
\textbf{Uplink} &&&&&&&&& \\
STUN & 124 & 1287.48 & 0.0008 & 124 & 1288.21 & 0.0008 & 124 & 1288.50 & 0.0008 \\
SRTCP & 434.62 & 67.27 & 0.052 & 421.68 & 63.97 & 0.053 & 418.92 & 63.56 & 0.053 \\
DTLS & 174.97 & 4.90 & 0.31 & 174.96 & 4.38 & 0.32 & 174.96 & 4.46 & 0.32 \\
Generic UDP & 147.64 & 443.75 & 0.0026 & 143.31 & 576.86 & 0.0020 & 138.96 & 639.26 & 0.0017 \\
\bottomrule
\end{tabular}
\end{table*}

Table \ref{tab:traffic_streams_characteristics} provides a comprehensive overview of the characteristics of each generated stream. Notably, the DL constitutes $\approx 99.3\%$ of the traffic load. The SRTP video stream accounts for the largest volume of data in the DL ($\approx 99.5\%$), while DTLS packets are the primary traffic in the UL ($\approx 85.0\%$). 

The different traffic streams exhibit similar traffic characteristics across distinct games. However, the Escape Room’s SRTP Audio stream displays a significantly higher load due to the presence of an active background audio source, distinguishing it from the other games that only include sound effects when certain actions are played. Additionally, UL generic UDP packets, which are deemed insignificant with respect the total traffic, show variations in packet timing between the three games.

Therefore, building upon the shared characteristics observed across games, the rest of the paper relies on Alteration Hunting as a representative sample.

%------------------------------------
%------------------------------------
\subsection{Temporal patterns}

\begin{figure}[t!!]
  \centering
  \includegraphics[width=\columnwidth]{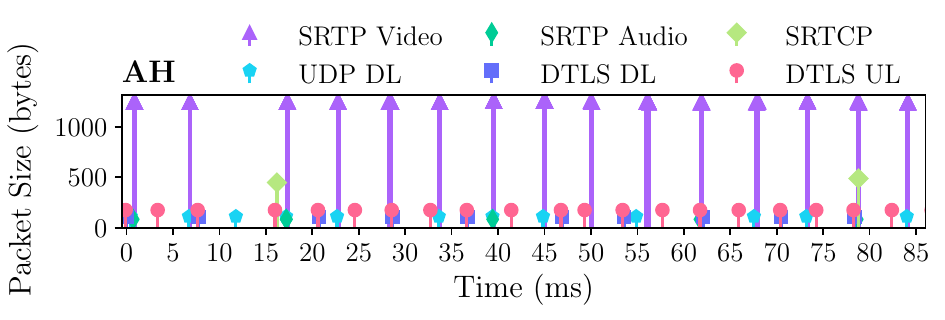}
  \caption{Traffic streams temporal evolution at 90 fps.}
  \label{fig:Streams_Temporal_Evolution}
\end{figure}

The SRTP Video stream exhibits periodic bursts of data, surprisingly surpassing the anticipated intervals according to the 90 fps target frame rate, as illustrated in Fig.~\ref{fig:Streams_Temporal_Evolution}. Given this unexpected phenomenon, and the fact that the video stream constitutes more than 98\% of the traffic, we devote Section~\ref{sec:video_traffic} to studying its behavior and characteristics.

\begin{figure}[tt!]
  \centering
  \includegraphics[width=\columnwidth]{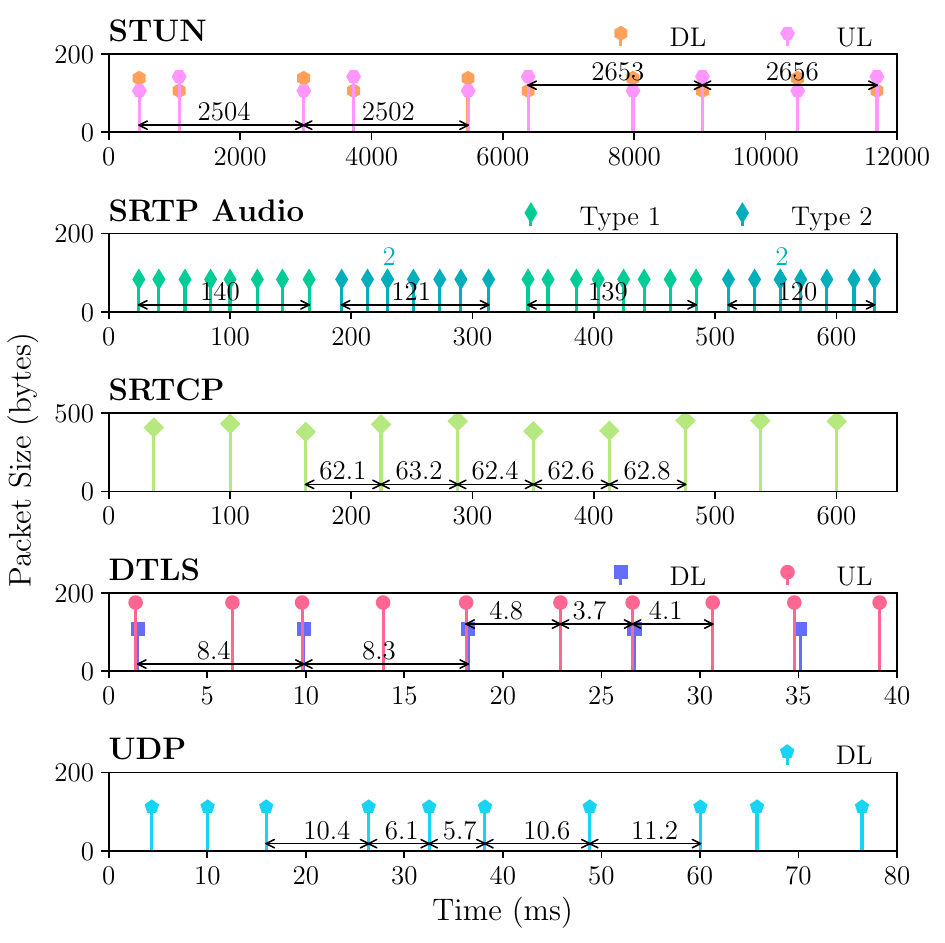}
  \caption{Traffic pattern of each stream.}
  \label{fig:Periodic_Streams}
\end{figure}

Fig.~\ref{fig:Periodic_Streams} shows that the remaining streams demonstrate discernible temporal patterns:
\begin{itemize}
    \item STUN traffic displays the recurring exchange of binding requests and 106-byte responses. DL requests and their responses occur every $\approx$ 2.5~s, while UL requests and their responses happen every $\approx$ 2.65~s.
    \item The SRTP Audio traffic generation pattern discerns the systematic arrangement of packets into alternating batch types, even in the absence of an active audio source. These batches comprise 8 and 7 arrival instances, spanning $\approx$ 140~ms and $\approx$ 120~ms, respectively, with a $\approx$ 30~ms gap between successive batches.
    \item SRTCP packets are transmitted at regular $\approx$ 62.5~ms intervals, regardless of the frame rate.
    \item UL DTLS packets are generated at the client’s display refresh rate (240~Hz, every $\approx$ 4.16~ms), while DL DTLS packets occur at half the frequency (120~Hz, every $\approx$ 8.33~ms).
    \item Generic UDP packets exhibit a bimodal distribution of inter-packet times in the DL (every $\approx$ 5.56~ms and $\approx$ 11.1~ms), while no discernible pattern is observed in the UL.
\end{itemize}

%------------------------------------
%------------------------------------
%------------------------------------
%------------------------------------

\section{Video traffic}\label{sec:video_traffic}

This section studies the video stream traffic, placing the focus on the occurrence of video traffic batches that surpass the targeted frame rate.
The analysis encompasses server-side Alteration Hunting traces at several target frame rates, including 90, 60, and 30~fps.

%------------------------------------
%------------------------------------
\subsection{Characteristics of video packets}

\begin{figure}[t]%h
  \centering
  \subfloat[]{\includegraphics[width=0.7\columnwidth]{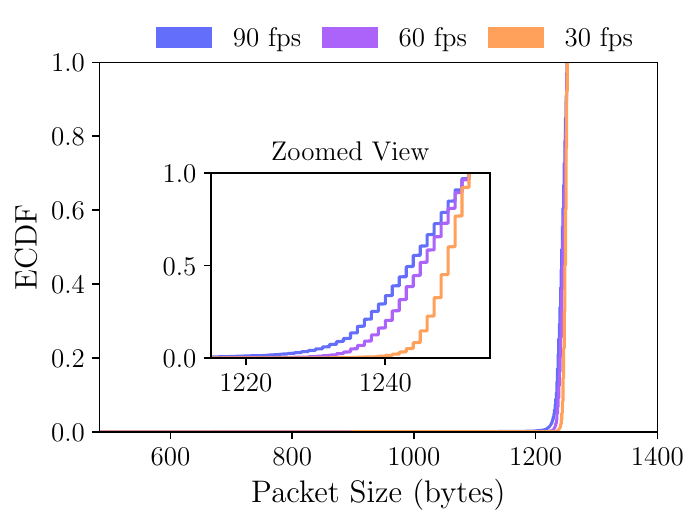}%
  \label{fig:Video_Packet_Size_ECDF_FPS}}
  \vfil
  \subfloat[]{\includegraphics[width=0.7\columnwidth]{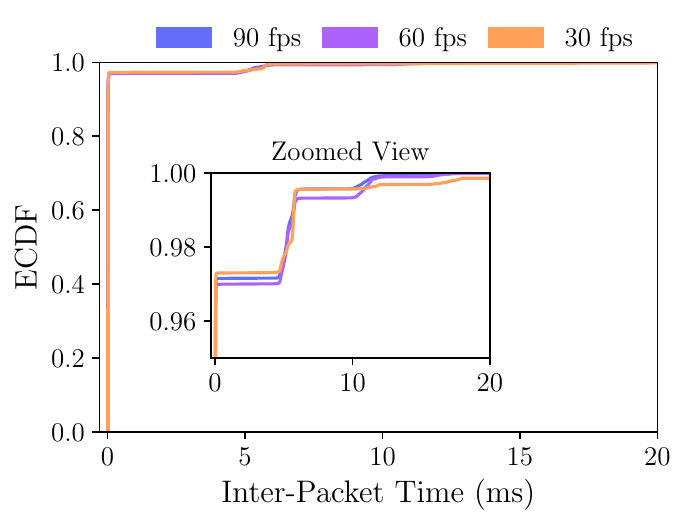}%
  \label{fig:Inter_Packet_Time_ECDF_FPS}}
  \caption{Video packet size ECDF (a) and inter-packet time ECDF (b).}
  \label{fig:Video_Packets_Characteristics}
\end{figure}

The SRTP Video stream exhibits consistent packet characteristics across games, as shown in Table~\ref{tab:traffic_streams_characteristics}. Notably, as depicted in Fig.~\ref{fig:Video_Packet_Size_ECDF_FPS}, video packet sizes strongly skew towards the 1220 to 1252~Bytes range. On the other hand, Fig.~\ref{fig:Inter_Packet_Time_ECDF_FPS} illustrates that a substantial portion of packets ($\approx 97$\%) exhibit microsecond-level inter-packet times, while the remaining packets manifest intervals in the order of milliseconds, forming discernible batches. 

%------------------------------------
%------------------------------------
\subsection{Characteristics of video frames} \label{sec:vfs_charact}

\begin{figure}[t] %h
  \centering
  \subfloat[]{\includegraphics[width=0.695\columnwidth]{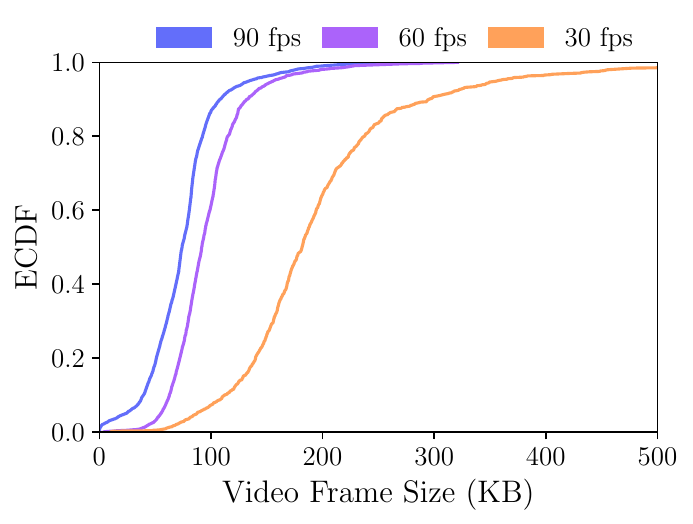}%
  \label{fig:Video_Frame_Size_ECDF_FPS}}
  \vfil
  \subfloat[]{\includegraphics[width=0.7\columnwidth]{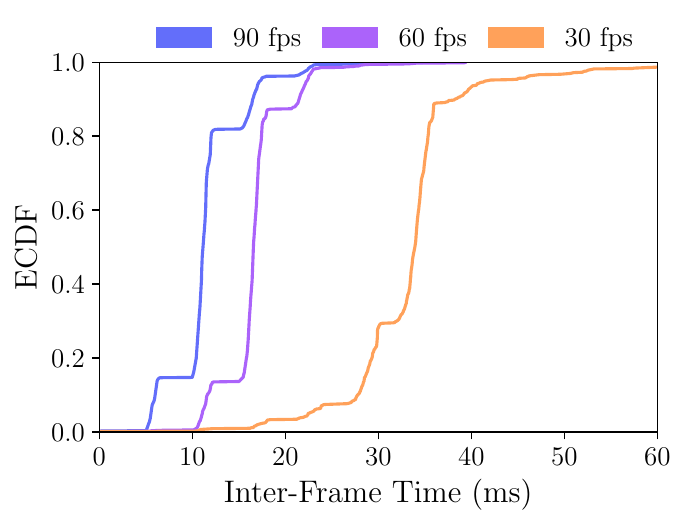}%
  \label{fig:Video_Inter_Frame_Time_ECDF_FPS}}
  \caption{VF size ECDF (a) and inter-frame time ECDF (b).}
  \label{fig:Video_Frames_Characteristics}
\end{figure}

Video content is transmitted as sequences of VFs with variable sizes. Intuitively, since the bitrate is fixed to 50~Mbps regardless of the frame rate, Fig.~\ref{fig:Video_Frame_Size_ECDF_FPS} shows that distributing similar data volumes over a higher number of frames leads to lower VF sizes ---averaging 201~KB at 30~fps, 99~KB at 60~fps, and 77~KB at 90~fps. Conversely, Fig.~\ref{fig:Video_Inter_Frame_Time_ECDF_FPS} illustrates the ECDF of the time lapse between consecutive VFs. Consistent with our expectations, the inter-frame period is closely correlated with the frame rate, averaging around $1/\text{fps}$: 33.33~ms at 30~fps, 16.67~ms at 60~fps, and 11.11~ms at 90~fps. 

%------------------------------------
%------------------------------------
\subsection{Video batching}
\label{sec:video_batching}

\begin{figure}[tt!]
  \centering
  \includegraphics[width=\columnwidth]{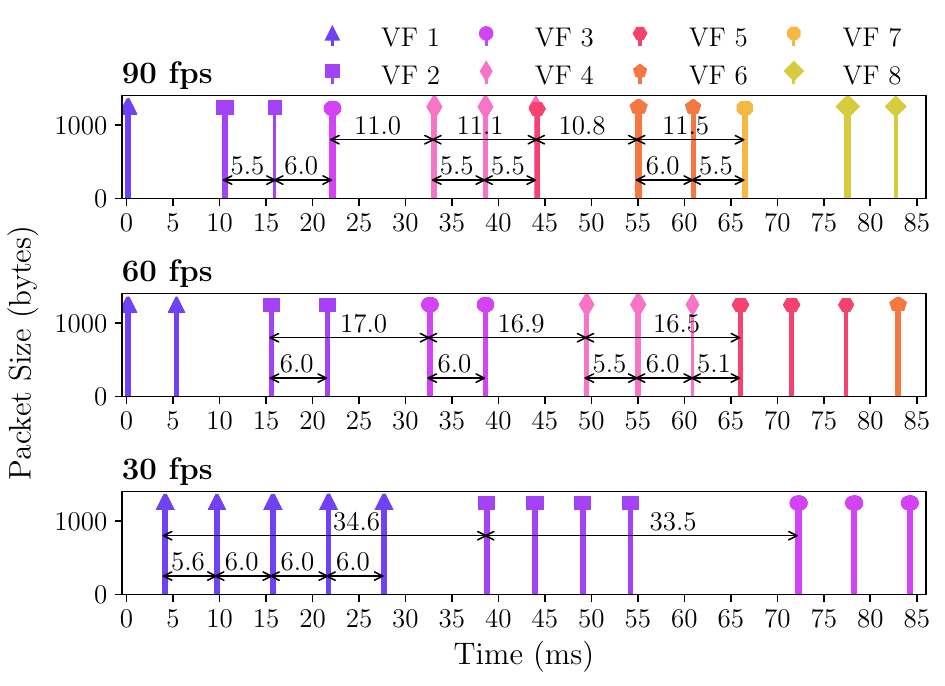}
  \caption{Video stream temporal evolution for Alteration Hunting at different fps.}
  \label{fig:Video_Stream_Temporal_Evolution}
\end{figure}

Fig.~\ref{fig:Video_Stream_Temporal_Evolution} shows that VFs are transmitted in batches of SRTP packets at regular intervals, regardless of the targeted frame rate. In particular, batch transmission points occur every $\approx 5.56$~ms.  
Nevertheless, idle instances arise when a complete frame is dispatched before the initiation of the subsequent frame's transmission. 

This transmission approach might be attributed to the \textit{paced sender} (or \textit{pacer}) component within Google’s WebRTC RTP stack~\cite{webrtc_pacing}.
Indeed, the GCC algorithm Internet-Draft recommends a 5~ms interval for the pacer to send a burst of packets, closely resembling the video traffic pattern observed in our tests~\cite{website:rfcDraftGCC}.
Once media RTP packets are queued in the pacer's buffer, the pacer regulates their transmission rate onto the network to mitigate potential network congestion, saturation, and packet loss that may arise from sending numerous UDP packets in a single burst.
Thus, the pacer sequentially releases a burst of video packets from its buffer for transmission, irrespective of their VF. This leads to batches encompassing packets from distinct VFs. For instance, in Fig.~\ref{fig:Video_Stream_Temporal_Evolution} at 90~fps, the batch transmitted at 45~ms includes packets from VFs~4 and~5.

\begin{figure}[th]
  \centering
  \includegraphics[width=0.7\columnwidth]
  {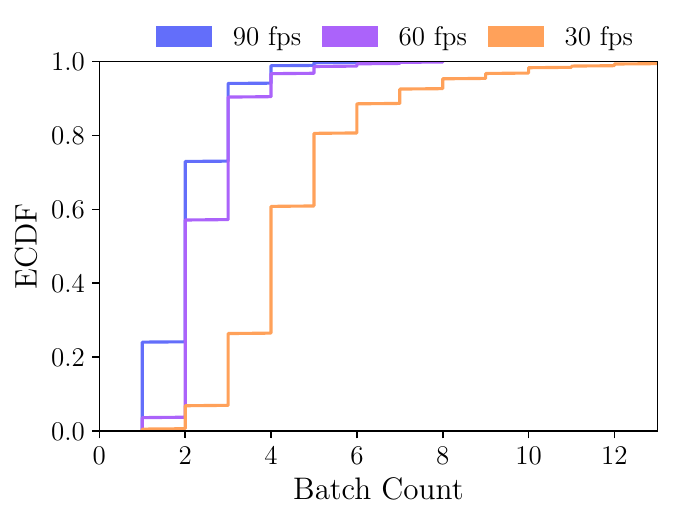}
    \caption{ECDF of the number of batches used for transmitting a VF.}
  \label{fig:Batch_count_per_frame_ECDF_FPS}
\end{figure}

\begin{figure}[th]
  \centering
  \includegraphics[width=0.7\columnwidth]{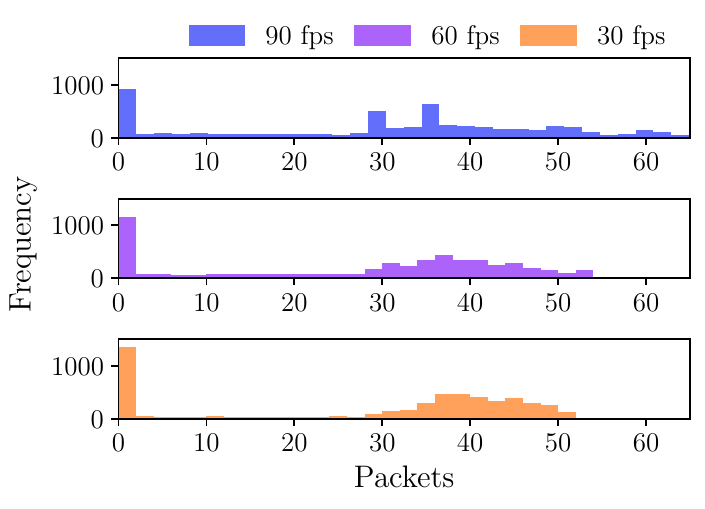}
    \caption{Histogram of video packets sent every 5.56 ms.}
  \label{fig:Histogram_of_Video_Packets_sent_every_5_56ms}
\end{figure}

At lower frame rates, VFs exhibit heightened packetization due to their larger sizes, as illustrated in Fig.~\ref{fig:Video_Frame_Size_ECDF_FPS}. Additionally, lower frame rates entail longer intervals between frames, as shown in Fig.~\ref{fig:Video_Inter_Frame_Time_ECDF_FPS}. Consequently, a higher number of batches are used for transmitting VFs at lower frame rates, as depicted in Fig.~\ref{fig:Batch_count_per_frame_ECDF_FPS}. Specifically, the average number of batches for transmitting VFs is 2.1, 2.54, and 4.56 batches at 90, 60, and 30~fps, respectively.
Nevertheless, prolonged inter-frame periods also result in additional transmission points, leading to more idle transfer intervals. Indeed, Fig.~\ref{fig:Histogram_of_Video_Packets_sent_every_5_56ms} reveals that around 15\% of the transfer instances occurring every $5.56$~ms exhibit no packets at 90~fps. In contrast, idle instances rise to 20\% and 25\% at 60 and 30~fps, respectively.

%------------------------------------
%------------------------------------
%------------------------------------
%------------------------------------
\section{Streaming performance}\label{sec:streaming_performance}

In this section, we shift our focus towards validating the streaming performance and QoS attainment, leveraging WebRTC statistics and traffic traces gathered from both ends during the streaming of Alteration Hunting. Specifically, encoded fps and RTT source from server-side WebRTC statistics, whereas decoded fps and jitter originate from client-side WebRTC statistics. On the other hand, assembly delays derive from client-side traffic traces.

\subsection{Decoded fps}

The number of frames decoded per second provides valuable insights into the stability and consistency of the streaming system.

\begin{figure}[th]
  \centering
  \includegraphics[width=\columnwidth]{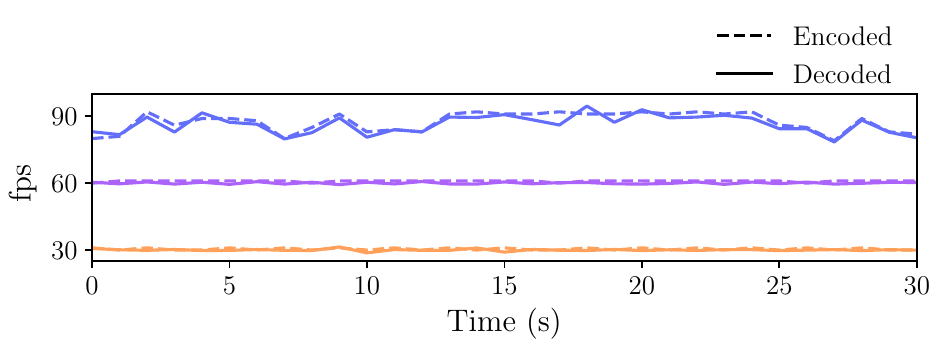}
  \caption{Encoded and decoded fps temporal evolution.}
  \label{fig:Encoded_and_Decoded_FPS_Temporal_Evolution}
\end{figure}

Notably, Fig.~\ref{fig:Encoded_and_Decoded_FPS_Temporal_Evolution} confirms our system's reliable operation up to 90~fps, given that encoded frames are successfully decoded and the target frame rate is nearly attained. Interestingly, at 90~fps, Unity Render Streaming appears to approach its operational threshold as the encoder struggles to maintain a consistent video frame rate. Nevertheless, the gaming experience at 90~fps displayed enhanced smoothness, responsiveness, and fluidity. On the other hand, there were no discernible differences in terms of perceptual quality between 90, 60, and 30~fps.

\begin{figure*}[t!!]
  \centering
  \begin{subfigure}[b]{0.3\linewidth}
    \includegraphics[width=0.98\linewidth]{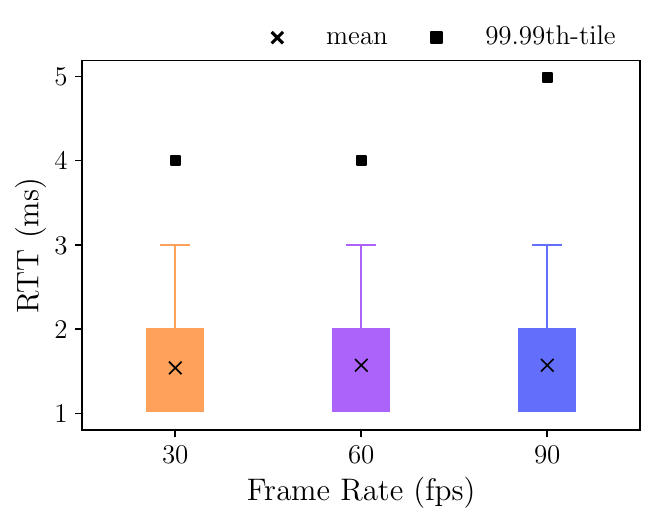}
    \caption{}
    \label{fig:RTT}
  \end{subfigure}
  \begin{subfigure}[b]{0.3\linewidth}
    \includegraphics[width=\linewidth]{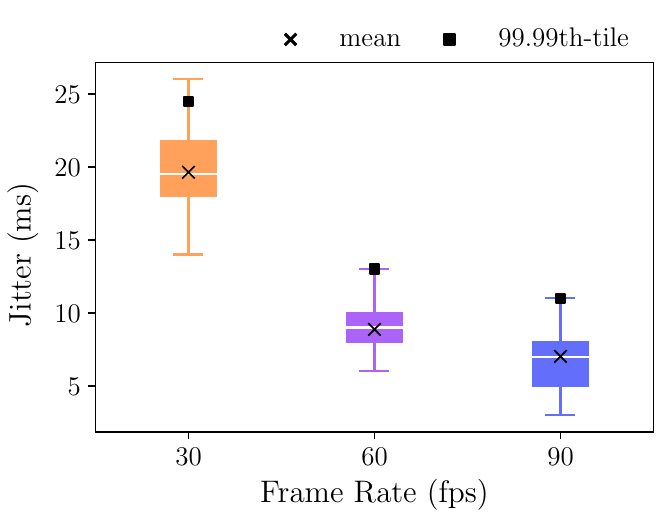}
    \caption{}
    \label{fig:Jitter}
  \end{subfigure}
  \begin{subfigure}[b]{0.3\linewidth}
    \includegraphics[width=\linewidth]{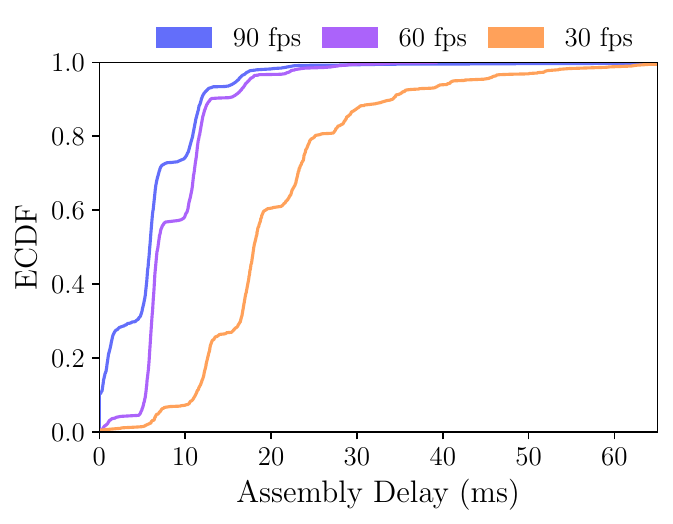}
    \caption{}
    \label{fig:Assembly_Delays_FPS}
  \end{subfigure}
  \caption{Streaming performance metrics at different fps.}
  \label{fig:streaming_performance}
\end{figure*}

\subsection{Key performance indicators}

RTT, jitter, and packet loss are crucial indicators of network performance and key factors for assessing QoS in WebRTC-based services \cite{garcia2019understanding}. According to the International Telecommunication Union Telecommunication Standardization Sector (ITU-T), VR services require RTT values below 20~ms, jitter levels under 15~ms, and a packet loss rate beneath 0.001\% \cite{itu_t_y3109, itu_t_j1631}.

Interestingly, Fig.~\ref{fig:RTT} displays uniform RTT performance across frame rates, averaging 1.5~ms.
Notably, the 99.99th percentile values, denoting the `worst-case' observed latency, stand at 4~ms for 30 and 60~fps, and at 5~ms for 90~fps.
Thus, the reported RTT values, estimated based on RTCP timestamps, align with ITU-T recommendations, contributing to motion sickness prevention.  
Fig.~\ref{fig:Jitter} shows that higher frame rates lead to reduced video jitter delay. Specifically, per-packet video jitter values revolve around 20~ms for 30~fps, 9~ms for 60~fps, and 7~ms for 90~fps, consistently adhering to ITU-T guidelines for 60~fps and 90~fps. This increase in jitter at lower frame rates may stem from larger VF sizes and heightened idle batch transfer instances, contributing to a less evenly spaced delivery of packets.

Throughout each trial, a zero packet loss ratio was consistently achieved, facilitated by favorable experimental conditions and the efficacy of WebRTC's retransmission mechanism, as evidenced by the presence of retransmissions in the statistics\footnote{Retransmissions during our tests represented less than 0.03\% of the total packet transmissions}. Thus, our VR streaming system successfully met the QoS goals at target frame rates of 60 and 90~fps.

\subsection{Assembly delay}
\label{sec:assembly_delay}

Assembly delay refers to the time elapsed between the reception of the first and last SRTP packet within a VF. Hence, minimizing this temporal gap is paramount for the timely delivery of seamless and responsive experiences.

Interestingly, as illustrated in Fig.~\ref{fig:Assembly_Delays_FPS}, lower frame rates exhibit larger assembly delays. Specifically, assembly delays average around 21~ms at 30~fps, 9~ms at 60~fps, and 6~ms at 90~fps.
This is a result of pacing video packet transmissions at 5.56~ms intervals. Indeed, this transmission mechanism inherently leads to longer assembly delays than delivering a VF in shorter intervals or in a single batch.

%------------------------------------
%------------------------------------
%------------------------------------
%------------------------------------
\section{Wi-Fi and VR traffic interplay}
\label{sec:wifi_vr_traffic_interplay}

In the previous sections, we delved into the intricacies of the generated traffic and validated the streaming performance.
Nevertheless, the wireless network infrastructure can significantly disrupt VR traffic patterns, potentially compromising the quality of experience. 
In addition, the effects of delivering VFs in temporally spaced batches require closer scrutiny.
Hence, this section delves into the interplay between Wi-Fi networks and Unity’s WebRTC-based VR traffic, relying on both empirical data and simulations.

%------------------------------------
%------------------------------------

\subsection{Wi-Fi’s impact on VR traffic: real-world tests}

\begin{figure}[th!!]
  \centering
  \includegraphics[width=\columnwidth]{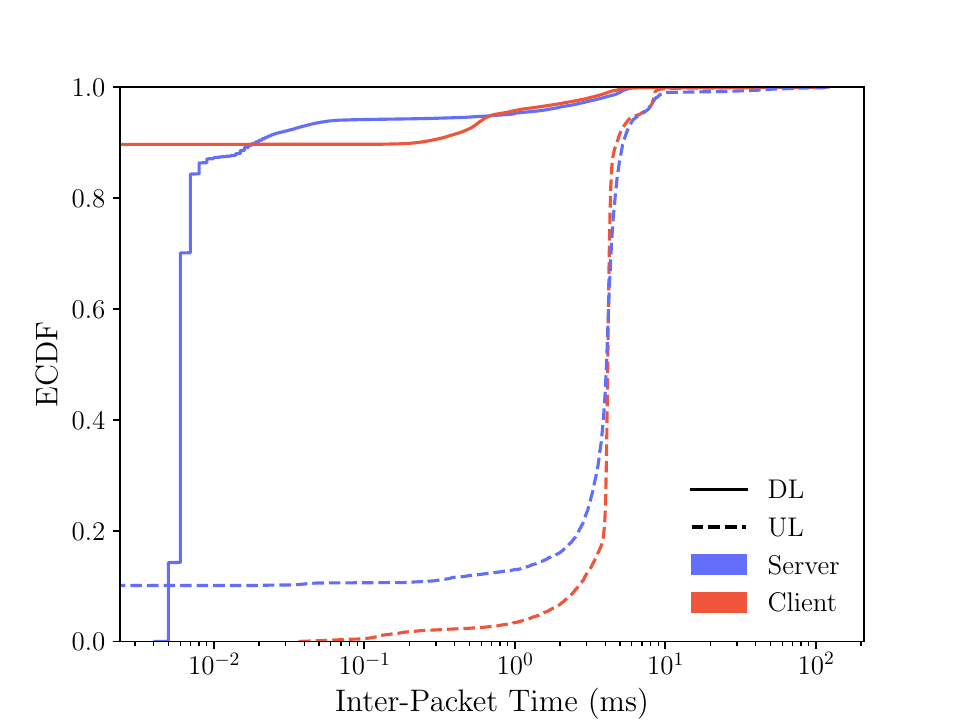}
  \caption{Inter-packet time ECDF of both DL and UL traffic.}
  \label{fig:Inter_Packet_Time_ECDF_DL_UL}
\end{figure}

To assess the network's impact on VR traffic dynamics, we first leverage Alteration Hunting traffic traces at 90~fps from both endpoints.
Fig.~\ref{fig:Inter_Packet_Time_ECDF_DL_UL} shows that the network significantly disrupts the timing of both DL and UL packets, showcasing the occurrence of simultaneous packet arrivals. In particular, 89.7\% of the DL packets and 10.1\% of the UL packets are received simultaneously. 

\begin{figure}[th!!!]
  \centering
  \includegraphics[width=\columnwidth]{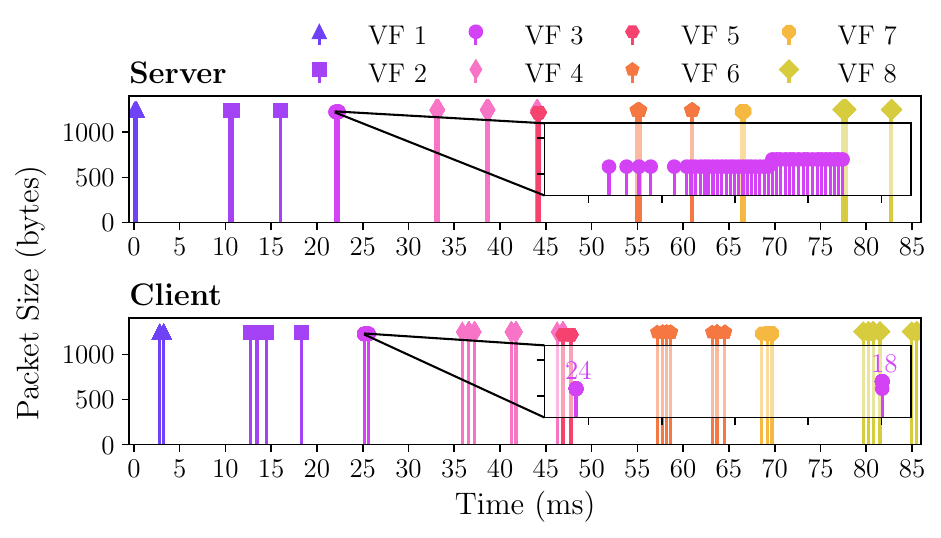}
  \caption{Video packet aggregation at 90 fps.}
  \label{fig:Video_Packet_Aggregation}
\end{figure}

This phenomenon stems from the accumulation of packets in the AP and station transmission buffers during channel access attempts, causing their aggregation into A-MPDUs, as depicted in Fig.~\ref{fig:Video_Packet_Aggregation} for DL video packets.

\begin{figure}[tt!]
  \centering
  \includegraphics[width=0.7\columnwidth]{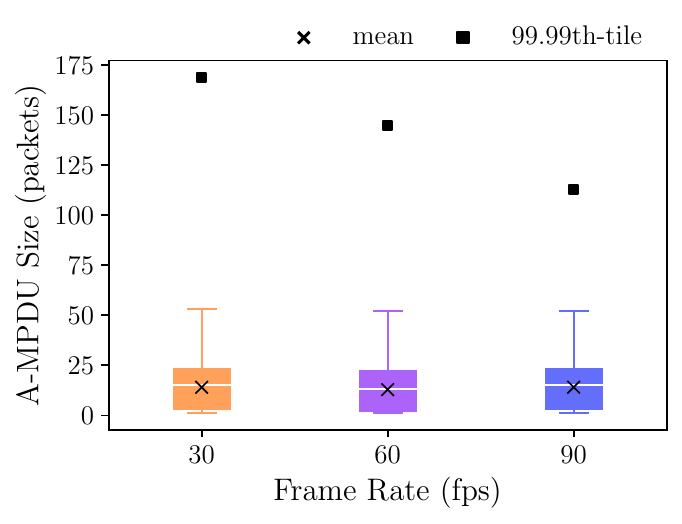}
  \caption{Video A-MPDU size for different fps: real-world tests.}
  \label{fig:ampdu_size_tests}
\end{figure}

Using client-side DL Alteration Hunting traces at 30, 60, and 90~fps, Fig.~\ref{fig:ampdu_size_tests} unveils consistent video A-MPDU sizes across frame rates, averaging 14, 13, and 14 packets at 30, 60, and 90~fps, respectively. Thus suggesting a stable data aggregation mechanism.
Low video frame rates should offer more opportunities for transmitting larger A-MPDUs, as VFs are larger in size, thus achieving a more efficient use of spectrum resources. However, the transmission of VFs in multiple temporally spaced batches prevents the use of large A-MPDU transmissions, leading to a nearly uniform mean A-MPDU size across frame rates. Moreover, since the server sends packets from a VF sequentially, the AP may not receive all packets before initiating a new transmission after channel contention. As a result, the AP can only assemble the packets present in the transmission buffer at that specific moment into an A-MPDU.

%------------------------------------
%------------------------------------

\subsection{Wi-Fi’s response to VR traffic: simulations}

The preceding subsection has shown the impact of Wi-Fi random channel access and packet aggregation on the delivery of VR video packets from the AP to the client. However, comprehensively analyzing the intricate relationship between Wi-Fi and VR traffic based solely on network-layer traffic traces proves challenging. 

In this section, we turn to simulations to examine how the division of VFs into multiple temporally spaced batches affects Wi-Fi's response, particularly in terms of latency and spectrum utilization. We leverage a Wi-Fi simulator\footnote{The Wi-Fi simulator used in this paper is based on the IEEE 802.11ax simulator ---codenamed Komondor--- developed at the Wireless Networking research group \cite{barrachina2019komondor}. The Komondor Wi-Fi simulator is available at \url{https://github.com/wn-upf/Komondor}} that implements the 802.11ax channel model, PHY, and MAC layers, using the same settings adopted during our tests ---namely, an 80~MHz bandwidth channel, 2~SS, and MCS~11, as described in Section \ref{sec:experimental_setup}. Wi-Fi simulation parameters, such as Packet Error Rate (PER), maximum number of retransmissions per packet, and Contention Window (CW) range, are outlined in Table~\ref{tab:sim_parameters}.
\begin{table}[h]
    \centering
    \caption{Simulation parameters.}
    \label{tab:sim_parameters}
    \footnotesize
    \begin{tabular}{ll|ll}
        \toprule
         \textbf{Parameter} & \textbf{Value} & \textbf{Parameter} & \textbf{Value}  \\
         Band & 5 GHz & Channel Width & 80 MHz \\
%         Tx. Power & 23 dBm & CCA & - 82 dBm \\
         MCS & 11 & AIFS & 34 $\mu_s$ \\ 
         CW$_{\min}$ & 31 & CW$_{\max}$ & 1023 \\ 
        SS & 2 & max A-MPDU & 256 pkts\\
         max Packet rtxs. & 7 & PER & 0.1 \\         
         AP tx. buffer & 1000 pkts & Client tx. buffer & 150 pkts \\
         Sim. duration & 10 s & Sim. runs & 10 \\
         \bottomrule
    \end{tabular}
\end{table}

The simulator incorporates an interactive application model that generates both VR video traffic in the DL and controller messages in the UL, replicating real traces. 
This entails reproducing the packetization of VFs and their transmission in batches at $\tau$~ms intervals (e.g., $\tau=5.56$~ms). Moreover, the simulator does not account for WebRTC's error resilience mechanisms, such as retransmissions, given their negligible impact on our tests.
In detail, VR traffic generation is modeled as follows:
\begin{enumerate}
    \item VFs are generated every $T=\frac{1}{\rm fps}$ seconds. The size in bits of each VF is constant, computed as the ratio between the bitrate (BR) and the fps, i.e., $L_{\rm VF}=\frac{\text{BR}}{\rm fps}$. Thus, using $\text{BR}=50$~Mbps, the size of each VF in simulations is consistent with the mean VF size in the traces, as detailed in Section~\ref{sec:vfs_charact}.
    \item The number of batches in which each VF is divided, $N_{b}$, is determined uniformly at random in the range $\left[1,\left\lceil{\frac{T}{\tau}}\right\rceil\right]$, where $\left\lceil{\frac{T}{\tau}}\right\rceil$ corresponds to the maximum number of batches that can be allocated for transmitting a VF before the subsequent one.
    \item The number of packets transmitted per batch, $N_{\rm pb}$, is determined by $N_{\rm pb}=\lceil{\frac{L_{\rm b}}{L_{\rm p}}}\rceil$, where $L_{\rm b}$ represents the size of each batch in bits, given by $L_{\rm b}=\frac{L_{\rm VF}}{N_{b}}$, and $L_{\rm p}$ denotes the packet size in bits. Notably, a $5~\mu$s delay is added between the generation of two consecutive packets to characterize processing delays at the server.
\end{enumerate}
In alignment with our real-world tests, a bitrate ($\text{BR}$) of 50~Mbps ---hence, $L_{\rm VF}$ equal to 70~KBytes at 90~fps--- and a packet size ($L_{\rm p}$) of 1243~Bytes are considered. Additionally, given that Unity Render Streaming uses an infinite GOP length, avoiding sending I-frames unless for error recovery ---as recommended by NVIDIA for low-latency applications \cite{nvidia_video_codec_sdk}--- our simulator does not consider periodic I-frame transmissions.

Before proceeding, let us establish a benchmark to understand the DL packet delays obtained in this section. Considering MCS~11, 2~SS, and an 80~MHz channel, a single video packet of 1243~Bytes requires 0.374~ms to travel from the server to the client, including RTS/CTS exchange, AIFS, and SIFS intervals between the AP and the client, as illustrated in Fig.~\ref{fig:DIFS_SIFS_process}. This value represents the minimum achievable latency.

%------------------------------------
\subsubsection{Matching Unity's 5.56~ms video batching}

In this section, we study the case when $\tau=5.56$~ms, matching Unity's traffic generation, as described in Section \ref{sec:video_traffic}. Note that by considering frame rates of 30, 60, and 90~fps, $\tau=5.56$~ms allows for a maximum of 6, 3, and 2 batches for transmitting a VF before the subsequent one, respectively. This aligns with a substantial portion of the VF transmissions in the tests, as depicted in Fig.~\ref{fig:Batch_count_per_frame_ECDF_FPS}.

\begin{figure}
    \centering
    \includegraphics[width=0.8\columnwidth]{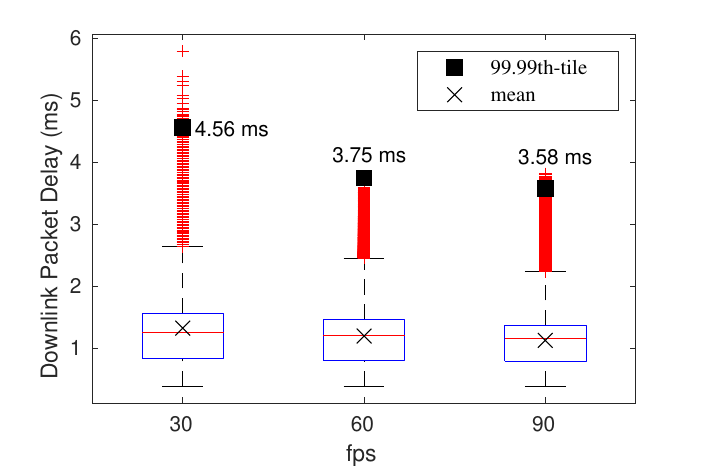}
    \caption{DL video packet delay for different fps.\vspace{3mm}}
    \label{fig:DL_vid_pkt_delay}
\end{figure}

\begin{figure}[t]
    \centering
    \includegraphics[width=0.8\columnwidth]{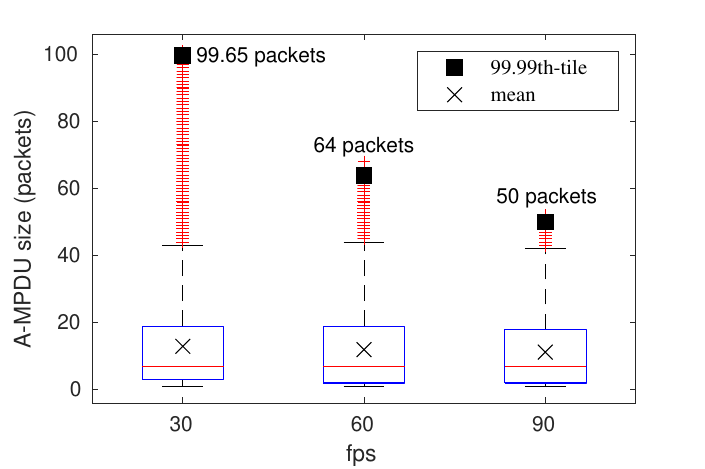}
    \caption{A-MPDU size for different fps.}
    \label{fig:UnityBufferAMPDU}
\end{figure}

Fig.~\ref{fig:DL_vid_pkt_delay} presents the DL delay in our simulations, including a boxplot representation, as well as the mean and 99.99th percentile values. The figure displays that the DL video packet delay distribution remains similar for different fps. Indeed, DL video packet delays slightly decrease as the fps increase, averaging 1.32~ms at 30~fps, 1.20~ms at 60~fps, and 1.12~ms at 90~fps. This subtle reduction in video packet delay may seem counter-intuitive, given that the airtime required to deliver all VR traffic rises from 16\% at 30~fps to 35\% at 90~fps, meaning that twice as many `spectrum resources' are required to transmit the same amount of traffic. Nevertheless, these results can be rationalized by the fact that the bitrate remains constant.
Thus, an increase in frame rate leads to smaller VFs and shorter intervals between VF transmissions, resulting in reduced buffer occupancy for incoming packets: from 26\% at 30~fps to 17\% at 90~fps. Reduced buffer occupancy contributes to a lower DL packet delay, as DL packet delay encompasses the entire time span a packet spends in the AP's transmission buffer, which depends on the number of packets present upon its arrival. However, lower buffer occupancy also implies that Wi-Fi aggregates fewer packets in each A-MPDU, as indicated by the 99.99th percentile values in Fig.~\ref{fig:UnityBufferAMPDU}. Hence, given that shorter A-MPDU transmissions are less efficient in terms of overhead reduction, more airtime is required to transmit the same volume of data. Notably, Fig.~\ref{fig:UnityBufferAMPDU} displays A-MPDU size distributions comparable to those in Fig.~\ref{fig:ampdu_size_tests}, with averages of 13, 12, and 11 packets at 30, 60, and 90~fps, respectively.

These results unveil the impact of transmitting VFs in batches at 5.56~ms intervals. This approach reduces the amount of data delivered to the network at every transfer instance, thus further minimizing the DL packet delay. Implementing this mechanism for every nominal frame rate maintains a nearly constant DL packet delay, although it comes at the cost of increased spectrum resource usage as Wi-Fi transmission becomes less efficient.

%------------------------------------

\subsubsection{Using distinct inter-batch times}

In this section, we investigate the influence of distinct inter-batch times ($\tau$) on DL video packet delays and VF delays to better understand the effects of delivering VFs in temporally spaced batches. VF delay refers to the time elapsed from the moment a VF is sent to the network interface at the server until it is completely received at the client, equivalent to the assembly delay.

\begin{figure}[ttt]
    \centering
    \includegraphics[width=0.8\columnwidth]{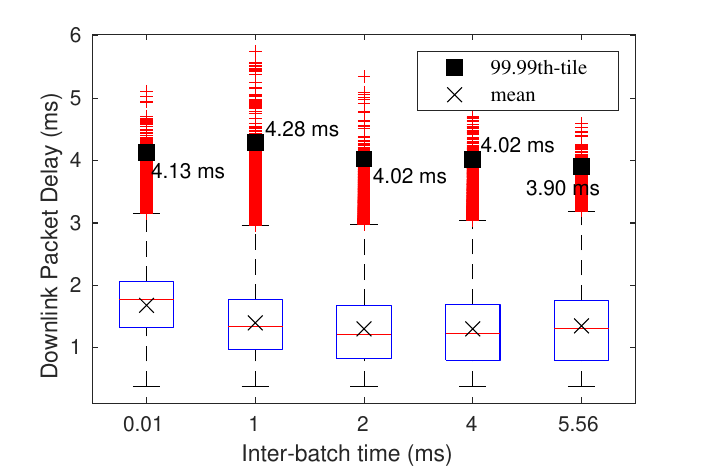}
    \caption{DL video packet delay at 60 fps for different inter-batch times.}
    \label{fig:UnityDLDelayVideoFrame_interbatch_1}
\end{figure}

\begin{figure}[ttt]
    \centering
    \includegraphics[width=0.8\columnwidth]{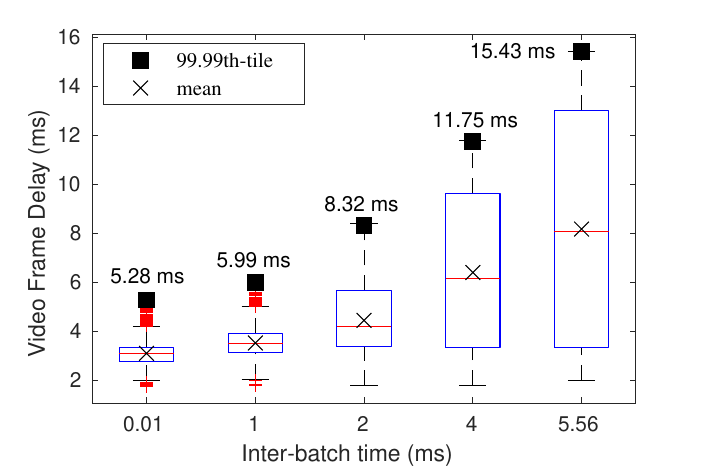}
    \caption{VF delay at 60 fps for different inter-batch times.}
    \label{fig:UnityDLDelayVideoFrame_interbatch_2}
\end{figure}

Figs.~\ref{fig:UnityDLDelayVideoFrame_interbatch_1} and~\ref{fig:UnityDLDelayVideoFrame_interbatch_2} show the DL video packet delay and VF delay for several inter-batch times at 60~fps. As it can be observed in Fig.~\ref{fig:UnityDLDelayVideoFrame_interbatch_1}, neither for the mean or the 99th percentile, increasing the inter-batch time has any significant effect beyond slightly reducing the mean packet delay.
Indeed, DL video packet delays average 1.68, 1.40, 1.30, 1.30, and 1.35~ms at inter-batch times of 0.01, 1, 2, 4, and 5.56~ms, respectively.
However, as depicted in Fig.~\ref{fig:UnityDLDelayVideoFrame_interbatch_2}, longer inter-batch times lead to prolonged VF delays. In particular, the average delays are 3.10, 3.52, 4.45, 6.40, and 8.18~ms for inter-batch times of 0.01, 1, 2, 4, and 5.56~ms, respectively.
Thus, since the interval between batches has minimal impact on DL packet delays, opting for a shorter inter-batch time is clearly the optimal choice from the Wi-Fi perspective.
Notably, the obtained mean VF delay using a 5.56~ms inter-batch time is consistent with the mean assembly delay in real-world tests at 60~fps outlined in Section \ref{sec:assembly_delay}.

%------------------------------------
%------------------------------------
%------------------------------------
%------------------------------------

\section{Conclusions}\label{sec:conclusions}

In this work, we empirically assessed Wi-Fi’s suitability for delivering remotely rendered VR gaming experiences, leveraging Unity.
Overall, the VR streaming system demonstrated sustained network performance and streaming quality at 60 and 90~fps.
Nevertheless, AP bundling of video packets and network congestion intervals led to disruptions in traffic patterns. 

In our in-depth traffic analysis, we discerned the existence of multiple periodic data streams, including UL tracking/input packets aligned with the client’s display refresh rate. Additionally, we identified frame rate aligned VF bursts within the SRTP video stream.
%, consistent with prior studies \cite{manzano2014dissecting, salehi2020traffic, carrascosa2022cloud, chiariotti2023temporal}. 
Notably, we discovered that VFs are delivered in multiple batches of SRTP packets transmitted every `5.56~ms', a pattern not unique to Unity Render Streaming but rather prevalent in WebRTC applications, as evidenced in \ref{sec:otherAPPS}.  
This mechanism introduces delay in the delivery of complete frames, potentially leading to increased latency and reduced responsiveness—critical aspects in real-time experiences. Additionally, while it allows to maintain a relatively consistent packet delay, it leads to increased Wi-Fi's airtime utilization at higher frame rates, given the aggregation of fewer packets per transmission.

This paper serves as a reference for future research in interactive VR cloud/edge gaming over Wi-Fi, building upon insights from edge-based VR streaming. It opens up new avenues for investigation, including the optimization of Unity’s video streaming parameters to enhance the delivery of remotely rendered VR experiences via Wi-Fi, as well as the evaluation of the system scalability and performance in dynamic network environments. Another intriguing research direction involves examining the intricate interplay between adaptive bitrate, WebRTC congestion mechanisms, and Wi-Fi features within the realm of VR cloud/edge gaming ---mathematically modeling the system operation and response whenever possible to better characterize the obtained findings. Studying the interaction between Wi-Fi and WebRTC error resilience mechanisms in VR streaming using a network emulator to introduce controlled packet errors and delays could also provide valuable insights. Moreover, using the insights gathered on Unity's Render Streaming VR traffic, more accurate VR traffic models can be developed ---incorporating key VR traffic features for modeling, such as VF size distributions, as indicated in~\cite{korneev2024model}. Furthermore, exploring the impact of background traffic or multiplayer VR streaming could shed light on how Wi-Fi manages multiple concurrent data streams.

%------------------------------------
%------------------------------------
%------------------------------------
%------------------------------------

\section{Acknowledgments}\label{ACKs}

This work is partially funded by Wi-XR PID2021-123995NB-I00 (MCIU/AEI/FEDER,UE), MAX-R EU-HE2022 (101070072), by MCIN/AEI under the Maria de Maeztu Units of Excellence Programme (CEX2021-001195-M) and 2021-SGR-00955. We would like to acknowledge the insightful feedback received from the Editor, Prof. Damla Turgut, and the anonymous Reviewers to enhance the final quality of this work.

%------------------------------------
%------------------------------------
%------------------------------------
%------------------------------------

\appendix

\section{VR traffic using a Meta Quest~2 HMD as the input source}\label{sec:HMDcomparative}
\begin{figure}[ht]
  \centering
\includegraphics[width=\columnwidth]{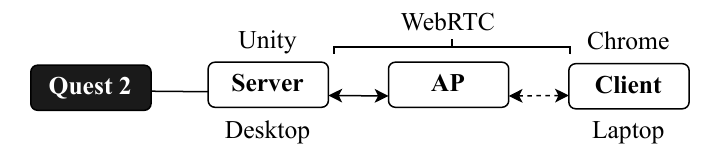}
  \caption{Testbed components using an HMD as the input source.}
  \label{fig:HMD_testbed_components}
\end{figure}

Is our video traffic characterization and streaming performance evaluation representative of scenarios involving HMDs?
As described in Section~\ref{subsec:unity_render_streaming}, Unity Render Streaming does not feature sending VR input messages from the web client to Unity nor mapping these remote inputs to Unity actions. 
In order to enable genuine VR input, we leveraged a Meta Quest~2 HMD and Meta Quest Link ---a feature that enables connecting a Meta Quest VR headset to a computer using a compatible USB-C cable, providing VR access to applications launched directly in Unity's Editor\footnote{\url{https://developer.oculus.com/documentation/unity/unity-link/}}.
Thus, an HMD served as the local input source, allowing us to broadcast realistic VR gameplay rendered on Unity from the server (desktop) to the client (laptop) via Unity Render Streaming, as illustrated in Fig.~\ref{fig:HMD_testbed_components}.

\begin{figure}[ht!!!]
  \centering
  \includegraphics[width=\columnwidth]{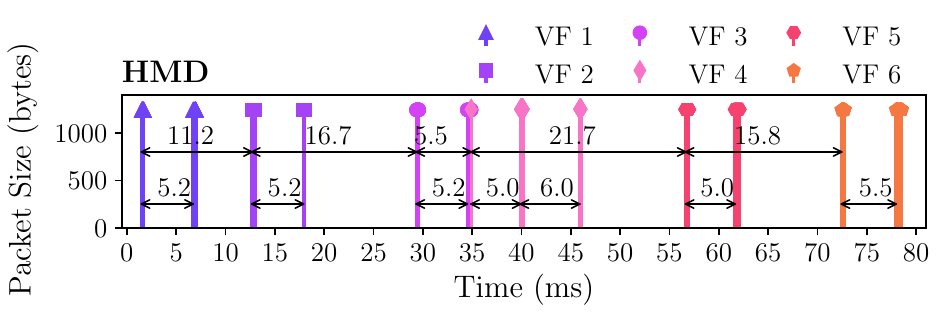}
  \caption{Video stream temporal evolution using Meta Quest Link and Alteration Hunting at 90~fps.}
  \label{fig:HMD_Temporal_Evolution}
\end{figure}

Notably, using Alteration Hunting and a 90~fps target frame rate, our analysis validates the system's performance when inputs are sourced from an HMD. In particular, no packet losses occurred during transmission, and the RTT and video jitter averaged 1.8~ms and 7.3~ms, respectively ---meeting VR services' QoS requirements. Additionally, the generated video stream remained consistent whether inputs originated from an HMD or a laptop.
Indeed, the only significant traffic disparity lied in the encoded frame rate, as Unity synchronizes with the HMD's default 72~Hz display refresh rate, as shown in Fig.~\ref{fig:HMD_Temporal_Evolution}. 

%------------------------------------
%------------------------------------
%------------------------------------
%------------------------------------
\begin{figure*}[tth!!]
  \centering    \includegraphics[width=0.85\linewidth]{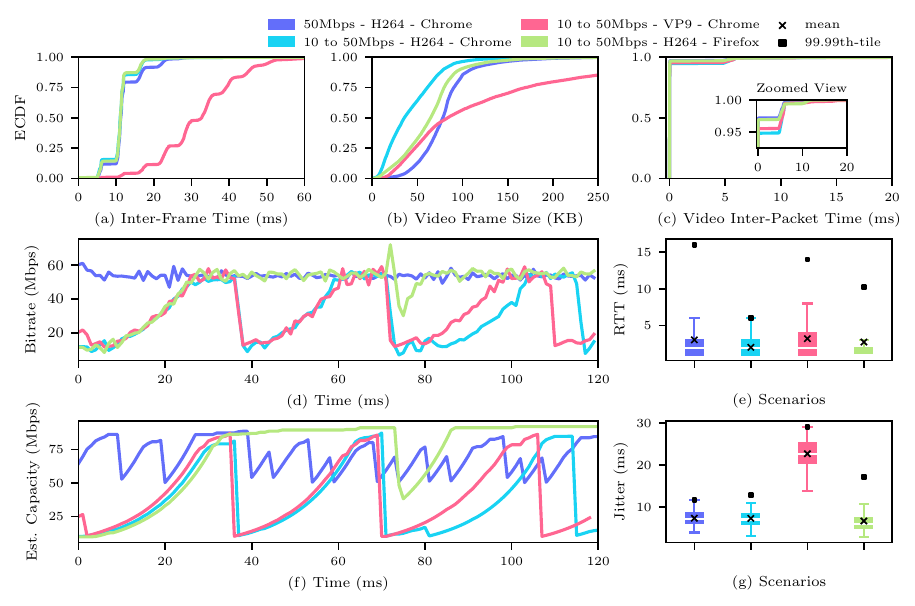}
    \caption{Constant bitrate and Dynamic bitrate scenarios using Unity Render Streaming.}
    \label{fig:VBR_tests_multifigure}
\end{figure*}

\section{Video streaming bitrate settings in Unity Render Streaming}\label{sec:CBRvsDBR}

This section investigates Unity’s Render Streaming video streaming bitrate parameters, contrasting constant bitrate scenarios, where both minimum and maximum bitrates are set to 50~Mbps, against dynamic bitrate scenarios, where the minimum is set to 10~Mbps and the maximum to 50~Mbps. The data collection encompasses server-side traffic traces and WebRTC statistics gathered over a 120-second Alteration Hunting active gameplay session, maintaining a target frame rate of 90~fps.

Let us begin by examining the scenarios employing the H.264 codec and Google Chrome browser, depicted in dark blue and light blue. As illustrated in Figs.~\ref{fig:VBR_tests_multifigure}a and ~\ref{fig:VBR_tests_multifigure}c, both scenarios demonstrate consistent video traffic patterns. VFs are transmitted every $1/\text{fps}$ on average in batches occurring at 5.56~ms intervals. Nevertheless, setting distinct minimum and maximum streaming bitrates leads to more variable frame size distributions, as evidenced in Fig.~\ref{fig:VBR_tests_multifigure}b.
In terms of QoS, both cases exhibit comparable performance. RTT and jitter distributions remain consistent, as shown in Figs.~\ref{fig:VBR_tests_multifigure}e and ~\ref{fig:VBR_tests_multifigure}g, respectively. The constant bitrate scenario averages an RTT of 3~ms and jitter of 7.2~ms, while the dynamic bitrate one averages an RTT of 2~ms and jitter of 7.2~ms. Surprisingly, using distinct minimum and maximum encoding bitrates introduces drops to the minimum bitrate, compromising the visual quality, as depicted in Fig.~\ref{fig:VBR_tests_multifigure}d. Notably, this adjustment is based on estimated capacity\footnote{\textit{availableOutgoingBitrate} field from \textit{RTCIceCandidatePairStats}}, as evidenced in Fig.~\ref{fig:VBR_tests_multifigure}f, rather than video content complexity. These drops to the minimum bitrate in dynamic bitrate scenarios persist regardless of the codec, as evidenced in Fig.~\ref{fig:VBR_tests_multifigure}d for the scenario using the VP9 software-based encoder, depicted in pink. Interestingly, using VP9, the encoding frame rates consistently fell below 30~fps, as evidenced in Fig.~\ref{fig:VBR_tests_multifigure}a. This observation highlights a struggle in meeting the targeted frame rate, indicating potential bottleneck issues. Indeed, VP9 exhibits elevated video jitter values averaging 22.6~ms, as shown in Fig.~\ref{fig:VBR_tests_multifigure}g. This level of jitter closely resembles the distribution observed using H.264 at 30~fps, depicted in Fig.~\ref{fig:Jitter}.
Interestingly, this phenomenon appears to be browser-specific, as evidenced in Fig.~\ref{fig:VBR_tests_multifigure}d when using Mozilla Firefox (v126.0), depicted in green. Using Firefox, the capacity estimate remains more consistent over time, as indicated in Fig.~\ref{fig:VBR_tests_multifigure}f. This finding is intriguing, especially considering that all scenarios employed the same Transport-Wide Congestion Control (TWCC) mechanism~\cite{website:draft_twcc} for bandwidth estimation. Thus, future endeavors may delve into browser-specific WebRTC implementations to elucidate the reason behind this behavior. Indeed, this phenomenon may be influenced by distinct congestion control implementations, such as GCC in Google Chrome and Network-Assisted Dynamic Adaptation (NADA)~\cite{website:RFC8698} in Mozilla Firefox.

%------------------------------------
%------------------------------------
%------------------------------------
%------------------------------------
\section{Video batching in other WebRTC-based applications}\label{sec:otherAPPS}

In Section \ref{sec:video_batching}, we unveiled Unity's Render Streaming plugin systematic partitioning of VFs into several batches of SRTP packets transmitted every 5.56~ms. This prompts a fundamental inquiry: Is the observed behavior caused by Unity's plugin?
To ascertain this, this section is devoted to investigating the traffic patterns of other WebRTC-based services, including GeForce NOW and Google Meet\footnote{\url{https://meet.google.com/}}. 

\begin{figure}[tt!]
  \centering
  \includegraphics[width=\columnwidth]{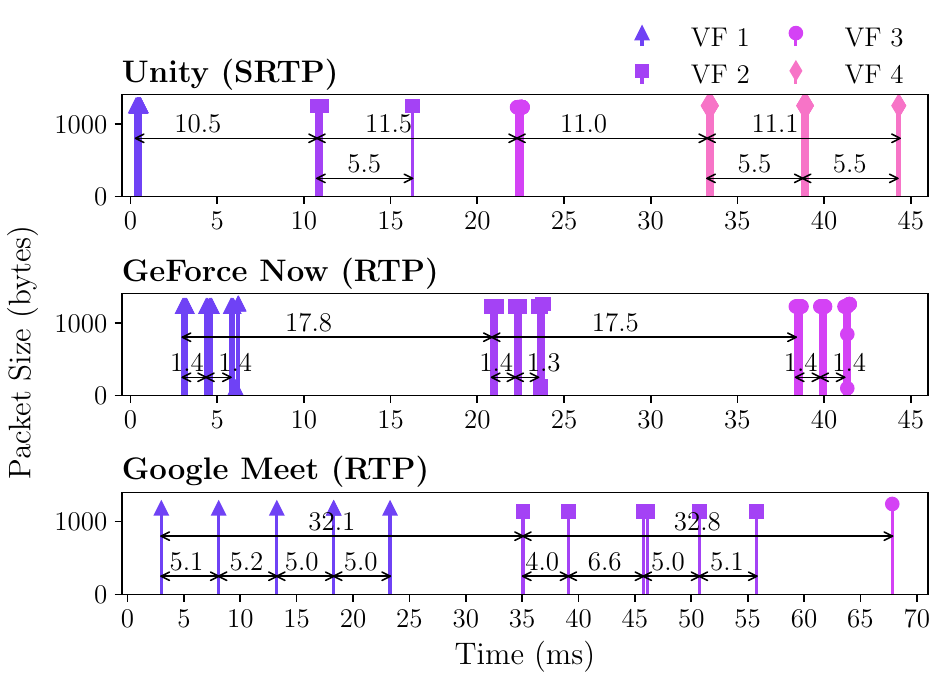}
  \caption{Video stream temporal evolution for several WebRTC-based services.} 
  \label{fig:Video_Batching_Comparative}
\end{figure}

As depicted in Fig.~\ref{fig:Video_Batching_Comparative}, GeForce NOW displays an inter-frame time averaging 16.68~ms, aligning with its free membership frame rate of 60~fps. On the other hand, Google Meet inter-frame time averages 33.43~ms, aligning with the broadcaster's 1280x720@30fps camera.
Interestingly, this figure underscores a prevalent pattern in WebRTC-based services, albeit with distinctions.
In particular, in GeForce NOW, VFs are delivered in several batches received at $\approx 1.4$ ~ms intervals.
Similarly, in Google Meet, VFs arrive in batches at intervals averaging 5.12~ms, consistent with Unity's Render Streaming behavior.
Hence, the transmission of video data in temporally spaced batches emerges as a consistent pattern in WebRTC-based services, each adopting a distinct approach. Indeed, in \cite{carrascosa2022cloud}, Google Stadia’s RTP stream also exhibited batches of video packets received every $\approx 2$~ms within each frame period.

%------------------------------------
%------------------------------------
%------------------------------------
%------------------------------------

\bibliographystyle{elsarticle-num} 
\bibliography{References}

\end{document}